\documentclass[longauth]{aa}

\usepackage{graphicx}
\usepackage{natbib}
\usepackage{scalerel}

\usepackage[table]{xcolor}

\usepackage{txfonts}
\usepackage[pdfencoding=auto,psdextra]{hyperref}
\hypersetup{
    colorlinks=true,
    linkcolor=blue,
    filecolor=magenta,      
    urlcolor=blue,
    citecolor=blue
}
\makeatletter
\renewcommand*\aa@pageof{, page \thepage{} of \pageref*{LastPage}}

\makeatother

\usepackage[utf8]{inputenc}

\usepackage[normalem]{ulem}
\usepackage{euclid}
\usepackage{preamble}

\newacronym{nisp}{NISP}{Near-Infrared Spectrometer and Photometer}
\newacronym{fov}{FOV}{field of view}
\newacronym{snr}{S/N}{signal-to-noise ratio}
\newacronym{ews}{EWS}{Euclid Wide Survey}
\newacronym{fs}{FS}{Flagship galaxy mock}
\newacronym{psf}{PSF}{Point Spread Function}
\newacronym{dr1}{DR1}{Data Release 1}
\newacronym{mad}{MAD}{Median Absolute Deviation}
\newacronym{spring}{\texttt{SPRING}}{Spectroscopic Pipeline Runner and INput Generator}
\newacronym{ial}{IAL}{Infrastructure Abstraction Layer}

\begin{document}

\newcommand{\orcid}[1]{} 
\author{Euclid Collaboration: F.~Passalacqua\orcid{0000-0002-8606-4093}\thanks{\email{francesca.passalacqua@pd.infn.it}}\inst{\ref{aff1},\ref{aff2}}
\and S.~Anselmi\orcid{0000-0002-3579-9583}\inst{\ref{aff2},\ref{aff1},\ref{aff3}}
\and P.~Monaco\orcid{0000-0003-2083-7564}\inst{\ref{aff4},\ref{aff5},\ref{aff6},\ref{aff7}}
\and C.~Sirignano\orcid{0000-0002-0995-7146}\inst{\ref{aff1},\ref{aff2}}
\and S.~Dusini\orcid{0000-0002-1128-0664}\inst{\ref{aff2}}
\and N.~Fourmanoit\orcid{0009-0005-6816-6925}\inst{\ref{aff8}}
\and M.~Fumana\orcid{0000-0001-6787-5950}\inst{\ref{aff9}}
\and E.~Lecrivain\orcid{0009-0003-5218-4278}\inst{\ref{aff8}}
\and K.~S.~McCarthy\orcid{0000-0001-6857-018X}\inst{\ref{aff10},\ref{aff11}}
\and M.~Moresco\orcid{0000-0002-7616-7136}\inst{\ref{aff12},\ref{aff13}}
\and F.~Oppizzi\orcid{0000-0003-3904-8370}\inst{\ref{aff14},\ref{aff2},\ref{aff1}}
\and A.~Renzi\orcid{0000-0001-9856-1970}\inst{\ref{aff1},\ref{aff2}}
\and M.~Scodeggio\inst{\ref{aff9}}
\and L.~Stanco\orcid{0000-0002-9706-5104}\inst{\ref{aff2}}
\and A.~Troja\orcid{0000-0003-0239-4595}\inst{\ref{aff1},\ref{aff2}}
\and S.~Bruton\orcid{0000-0002-6503-5218}\inst{\ref{aff15}}
\and C.~Carbone\orcid{0000-0003-0125-3563}\inst{\ref{aff9}}
\and S.~de~la~Torre\inst{\ref{aff16}}
\and B.~R.~Granett\orcid{0000-0003-2694-9284}\inst{\ref{aff17}}
\and G.~Lavaux\orcid{0000-0003-0143-8891}\inst{\ref{aff18}}
\and S.~Lee\orcid{0000-0002-8289-740X}\inst{\ref{aff10},\ref{aff19}}
\and K.~Markovic\orcid{0000-0001-6764-073X}\inst{\ref{aff10}}
\and W.~J.~Percival\orcid{0000-0002-0644-5727}\inst{\ref{aff20},\ref{aff21},\ref{aff22}}
\and I.~Risso\orcid{0000-0003-2525-7761}\inst{\ref{aff17},\ref{aff14}}
\and C.~Scarlata\orcid{0000-0002-9136-8876}\inst{\ref{aff23}}
\and E.~Sefusatti\orcid{0000-0003-0473-1567}\inst{\ref{aff5},\ref{aff7},\ref{aff6}}
\and Y.~Wang\orcid{0000-0002-4749-2984}\inst{\ref{aff24}}
\and S.~Andreon\orcid{0000-0002-2041-8784}\inst{\ref{aff17}}
\and N.~Auricchio\orcid{0000-0003-4444-8651}\inst{\ref{aff13}}
\and H.~Aussel\orcid{0000-0002-1371-5705}\inst{\ref{aff25}}
\and C.~Baccigalupi\orcid{0000-0002-8211-1630}\inst{\ref{aff7},\ref{aff5},\ref{aff6},\ref{aff26}}
\and M.~Baldi\orcid{0000-0003-4145-1943}\inst{\ref{aff27},\ref{aff13},\ref{aff28}}
\and S.~Bardelli\orcid{0000-0002-8900-0298}\inst{\ref{aff13}}
\and P.~Battaglia\orcid{0000-0002-7337-5909}\inst{\ref{aff13}}
\and A.~Biviano\orcid{0000-0002-0857-0732}\inst{\ref{aff5},\ref{aff7}}
\and E.~Branchini\orcid{0000-0002-0808-6908}\inst{\ref{aff29},\ref{aff14},\ref{aff17}}
\and M.~Brescia\orcid{0000-0001-9506-5680}\inst{\ref{aff30},\ref{aff31}}
\and S.~Camera\orcid{0000-0003-3399-3574}\inst{\ref{aff32},\ref{aff33},\ref{aff34}}
\and G.~Ca\~nas-Herrera\orcid{0000-0003-2796-2149}\inst{\ref{aff35},\ref{aff36}}
\and V.~Capobianco\orcid{0000-0002-3309-7692}\inst{\ref{aff34}}
\and V.~F.~Cardone\inst{\ref{aff37},\ref{aff38}}
\and J.~Carretero\orcid{0000-0002-3130-0204}\inst{\ref{aff39},\ref{aff40}}
\and S.~Casas\orcid{0000-0002-4751-5138}\inst{\ref{aff41},\ref{aff42}}
\and F.~J.~Castander\orcid{0000-0001-7316-4573}\inst{\ref{aff43},\ref{aff44}}
\and M.~Castellano\orcid{0000-0001-9875-8263}\inst{\ref{aff37}}
\and G.~Castignani\orcid{0000-0001-6831-0687}\inst{\ref{aff13}}
\and S.~Cavuoti\orcid{0000-0002-3787-4196}\inst{\ref{aff31},\ref{aff45}}
\and A.~Cimatti\inst{\ref{aff46}}
\and C.~Colodro-Conde\inst{\ref{aff47}}
\and G.~Congedo\orcid{0000-0003-2508-0046}\inst{\ref{aff48}}
\and C.~J.~Conselice\orcid{0000-0003-1949-7638}\inst{\ref{aff49}}
\and L.~Conversi\orcid{0000-0002-6710-8476}\inst{\ref{aff50},\ref{aff51}}
\and Y.~Copin\orcid{0000-0002-5317-7518}\inst{\ref{aff52}}
\and A.~Costille\inst{\ref{aff16}}
\and F.~Courbin\orcid{0000-0003-0758-6510}\inst{\ref{aff53},\ref{aff54},\ref{aff55}}
\and H.~M.~Courtois\orcid{0000-0003-0509-1776}\inst{\ref{aff56}}
\and A.~Da~Silva\orcid{0000-0002-6385-1609}\inst{\ref{aff57},\ref{aff58}}
\and H.~Degaudenzi\orcid{0000-0002-5887-6799}\inst{\ref{aff59}}
\and G.~De~Lucia\orcid{0000-0002-6220-9104}\inst{\ref{aff5}}
\and H.~Dole\orcid{0000-0002-9767-3839}\inst{\ref{aff60}}
\and F.~Dubath\orcid{0000-0002-6533-2810}\inst{\ref{aff59}}
\and C.~A.~J.~Duncan\orcid{0009-0003-3573-0791}\inst{\ref{aff48}}
\and X.~Dupac\inst{\ref{aff51}}
\and S.~Escoffier\orcid{0000-0002-2847-7498}\inst{\ref{aff8}}
\and M.~Farina\orcid{0000-0002-3089-7846}\inst{\ref{aff61}}
\and S.~Ferriol\inst{\ref{aff52}}
\and S.~Fotopoulou\orcid{0000-0002-9686-254X}\inst{\ref{aff62}}
\and M.~Frailis\orcid{0000-0002-7400-2135}\inst{\ref{aff5}}
\and E.~Franceschi\orcid{0000-0002-0585-6591}\inst{\ref{aff13}}
\and S.~Galeotta\orcid{0000-0002-3748-5115}\inst{\ref{aff5}}
\and K.~George\orcid{0000-0002-1734-8455}\inst{\ref{aff63}}
\and W.~Gillard\orcid{0000-0003-4744-9748}\inst{\ref{aff8}}
\and B.~Gillis\orcid{0000-0002-4478-1270}\inst{\ref{aff48}}
\and C.~Giocoli\orcid{0000-0002-9590-7961}\inst{\ref{aff13},\ref{aff28}}
\and J.~Gracia-Carpio\inst{\ref{aff64}}
\and A.~Grazian\orcid{0000-0002-5688-0663}\inst{\ref{aff65}}
\and F.~Grupp\inst{\ref{aff64},\ref{aff66}}
\and L.~Guzzo\orcid{0000-0001-8264-5192}\inst{\ref{aff67},\ref{aff17},\ref{aff68}}
\and S.~V.~H.~Haugan\orcid{0000-0001-9648-7260}\inst{\ref{aff69}}
\and W.~Holmes\inst{\ref{aff10}}
\and F.~Hormuth\inst{\ref{aff70}}
\and A.~Hornstrup\orcid{0000-0002-3363-0936}\inst{\ref{aff71},\ref{aff72}}
\and P.~Hudelot\inst{\ref{aff18}}
\and K.~Jahnke\orcid{0000-0003-3804-2137}\inst{\ref{aff73}}
\and M.~Jhabvala\inst{\ref{aff74}}
\and B.~Joachimi\orcid{0000-0001-7494-1303}\inst{\ref{aff75}}
\and E.~Keih\"anen\orcid{0000-0003-1804-7715}\inst{\ref{aff76}}
\and S.~Kermiche\orcid{0000-0002-0302-5735}\inst{\ref{aff8}}
\and A.~Kiessling\orcid{0000-0002-2590-1273}\inst{\ref{aff10}}
\and B.~Kubik\orcid{0009-0006-5823-4880}\inst{\ref{aff52}}
\and M.~Kunz\orcid{0000-0002-3052-7394}\inst{\ref{aff77}}
\and H.~Kurki-Suonio\orcid{0000-0002-4618-3063}\inst{\ref{aff78},\ref{aff79}}
\and A.~M.~C.~Le~Brun\orcid{0000-0002-0936-4594}\inst{\ref{aff80}}
\and S.~Ligori\orcid{0000-0003-4172-4606}\inst{\ref{aff34}}
\and P.~B.~Lilje\orcid{0000-0003-4324-7794}\inst{\ref{aff69}}
\and V.~Lindholm\orcid{0000-0003-2317-5471}\inst{\ref{aff78},\ref{aff79}}
\and I.~Lloro\orcid{0000-0001-5966-1434}\inst{\ref{aff81}}
\and G.~Mainetti\orcid{0000-0003-2384-2377}\inst{\ref{aff82}}
\and D.~Maino\inst{\ref{aff67},\ref{aff9},\ref{aff68}}
\and E.~Maiorano\orcid{0000-0003-2593-4355}\inst{\ref{aff13}}
\and O.~Mansutti\orcid{0000-0001-5758-4658}\inst{\ref{aff5}}
\and S.~Marcin\inst{\ref{aff83}}
\and O.~Marggraf\orcid{0000-0001-7242-3852}\inst{\ref{aff84}}
\and M.~Martinelli\orcid{0000-0002-6943-7732}\inst{\ref{aff37},\ref{aff38}}
\and N.~Martinet\orcid{0000-0003-2786-7790}\inst{\ref{aff16}}
\and F.~Marulli\orcid{0000-0002-8850-0303}\inst{\ref{aff12},\ref{aff13},\ref{aff28}}
\and R.~J.~Massey\orcid{0000-0002-6085-3780}\inst{\ref{aff85}}
\and E.~Medinaceli\orcid{0000-0002-4040-7783}\inst{\ref{aff13}}
\and S.~Mei\orcid{0000-0002-2849-559X}\inst{\ref{aff86},\ref{aff87}}
\and Y.~Mellier\thanks{Deceased}\inst{\ref{aff88},\ref{aff18}}
\and M.~Meneghetti\orcid{0000-0003-1225-7084}\inst{\ref{aff13},\ref{aff28}}
\and E.~Merlin\orcid{0000-0001-6870-8900}\inst{\ref{aff37}}
\and G.~Meylan\inst{\ref{aff89}}
\and A.~Mora\orcid{0000-0002-1922-8529}\inst{\ref{aff90}}
\and L.~Moscardini\orcid{0000-0002-3473-6716}\inst{\ref{aff12},\ref{aff13},\ref{aff28}}
\and E.~Munari\orcid{0000-0002-1751-5946}\inst{\ref{aff5},\ref{aff7}}
\and R.~Nakajima\orcid{0009-0009-1213-7040}\inst{\ref{aff84}}
\and C.~Neissner\orcid{0000-0001-8524-4968}\inst{\ref{aff91},\ref{aff40}}
\and R.~C.~Nichol\orcid{0000-0003-0939-6518}\inst{\ref{aff92}}
\and S.-M.~Niemi\orcid{0009-0005-0247-0086}\inst{\ref{aff35}}
\and C.~Padilla\orcid{0000-0001-7951-0166}\inst{\ref{aff91}}
\and S.~Paltani\orcid{0000-0002-8108-9179}\inst{\ref{aff59}}
\and F.~Pasian\orcid{0000-0002-4869-3227}\inst{\ref{aff5}}
\and K.~Pedersen\inst{\ref{aff93}}
\and V.~Pettorino\orcid{0000-0002-4203-9320}\inst{\ref{aff35}}
\and S.~Pires\orcid{0000-0002-0249-2104}\inst{\ref{aff25}}
\and G.~Polenta\orcid{0000-0003-4067-9196}\inst{\ref{aff94}}
\and M.~Poncet\inst{\ref{aff95}}
\and L.~A.~Popa\inst{\ref{aff96}}
\and L.~Pozzetti\orcid{0000-0001-7085-0412}\inst{\ref{aff13}}
\and G.~D.~Racca\orcid{0000-0002-9883-8981}\inst{\ref{aff35},\ref{aff36}}
\and F.~Raison\orcid{0000-0002-7819-6918}\inst{\ref{aff64}}
\and J.~Rhodes\orcid{0000-0002-4485-8549}\inst{\ref{aff10}}
\and G.~Riccio\inst{\ref{aff31}}
\and E.~Romelli\orcid{0000-0003-3069-9222}\inst{\ref{aff5}}
\and M.~Roncarelli\orcid{0000-0001-9587-7822}\inst{\ref{aff13}}
\and C.~Rosset\orcid{0000-0003-0286-2192}\inst{\ref{aff86}}
\and E.~Rossetti\orcid{0000-0003-0238-4047}\inst{\ref{aff27}}
\and R.~Saglia\orcid{0000-0003-0378-7032}\inst{\ref{aff66},\ref{aff64}}
\and Z.~Sakr\orcid{0000-0002-4823-3757}\inst{\ref{aff97},\ref{aff98},\ref{aff99}}
\and A.~G.~S\'anchez\orcid{0000-0003-1198-831X}\inst{\ref{aff64}}
\and D.~Sapone\orcid{0000-0001-7089-4503}\inst{\ref{aff100}}
\and B.~Sartoris\orcid{0000-0003-1337-5269}\inst{\ref{aff66},\ref{aff5}}
\and P.~Schneider\orcid{0000-0001-8561-2679}\inst{\ref{aff84}}
\and T.~Schrabback\orcid{0000-0002-6987-7834}\inst{\ref{aff101}}
\and A.~Secroun\orcid{0000-0003-0505-3710}\inst{\ref{aff8}}
\and G.~Seidel\orcid{0000-0003-2907-353X}\inst{\ref{aff73}}
\and S.~Serrano\orcid{0000-0002-0211-2861}\inst{\ref{aff44},\ref{aff102},\ref{aff43}}
\and P.~Simon\inst{\ref{aff84}}
\and G.~Sirri\orcid{0000-0003-2626-2853}\inst{\ref{aff28}}
\and J.~Steinwagner\orcid{0000-0001-7443-1047}\inst{\ref{aff64}}
\and C.~Surace\orcid{0000-0003-2592-0113}\inst{\ref{aff16}}
\and P.~Tallada-Cresp\'{i}\orcid{0000-0002-1336-8328}\inst{\ref{aff39},\ref{aff40}}
\and A.~N.~Taylor\inst{\ref{aff48}}
\and H.~I.~Teplitz\orcid{0000-0002-7064-5424}\inst{\ref{aff103}}
\and I.~Tereno\orcid{0000-0002-4537-6218}\inst{\ref{aff57},\ref{aff104}}
\and N.~Tessore\orcid{0000-0002-9696-7931}\inst{\ref{aff105}}
\and S.~Toft\orcid{0000-0003-3631-7176}\inst{\ref{aff106},\ref{aff107}}
\and R.~Toledo-Moreo\orcid{0000-0002-2997-4859}\inst{\ref{aff108}}
\and F.~Torradeflot\orcid{0000-0003-1160-1517}\inst{\ref{aff40},\ref{aff39}}
\and I.~Tutusaus\orcid{0000-0002-3199-0399}\inst{\ref{aff43},\ref{aff44},\ref{aff98}}
\and L.~Valenziano\orcid{0000-0002-1170-0104}\inst{\ref{aff13},\ref{aff109}}
\and J.~Valiviita\orcid{0000-0001-6225-3693}\inst{\ref{aff78},\ref{aff79}}
\and T.~Vassallo\orcid{0000-0001-6512-6358}\inst{\ref{aff5}}
\and A.~Veropalumbo\orcid{0000-0003-2387-1194}\inst{\ref{aff17},\ref{aff14},\ref{aff29}}
\and D.~Vibert\orcid{0009-0008-0607-631X}\inst{\ref{aff16}}
\and J.~Weller\orcid{0000-0002-8282-2010}\inst{\ref{aff66},\ref{aff64}}
\and G.~Zamorani\orcid{0000-0002-2318-301X}\inst{\ref{aff13}}
\and E.~Zucca\orcid{0000-0002-5845-8132}\inst{\ref{aff13}}
\and V.~Allevato\orcid{0000-0001-7232-5152}\inst{\ref{aff31}}
\and M.~Ballardini\orcid{0000-0003-4481-3559}\inst{\ref{aff110},\ref{aff111},\ref{aff13}}
\and M.~Bolzonella\orcid{0000-0003-3278-4607}\inst{\ref{aff13}}
\and A.~Boucaud\orcid{0000-0001-7387-2633}\inst{\ref{aff86}}
\and C.~Burigana\orcid{0000-0002-3005-5796}\inst{\ref{aff112},\ref{aff109}}
\and R.~Cabanac\orcid{0000-0001-6679-2600}\inst{\ref{aff98}}
\and M.~Calabrese\orcid{0000-0002-2637-2422}\inst{\ref{aff113},\ref{aff9}}
\and A.~Cappi\inst{\ref{aff114},\ref{aff13}}
\and J.~A.~Escartin~Vigo\inst{\ref{aff64}}
\and L.~Gabarra\orcid{0000-0002-8486-8856}\inst{\ref{aff115}}
\and W.~G.~Hartley\inst{\ref{aff59}}
\and R.~Maoli\orcid{0000-0002-6065-3025}\inst{\ref{aff116},\ref{aff37}}
\and J.~Mart\'{i}n-Fleitas\orcid{0000-0002-8594-569X}\inst{\ref{aff117}}
\and S.~Matthew\orcid{0000-0001-8448-1697}\inst{\ref{aff48}}
\and N.~Mauri\orcid{0000-0001-8196-1548}\inst{\ref{aff46},\ref{aff28}}
\and R.~B.~Metcalf\orcid{0000-0003-3167-2574}\inst{\ref{aff12},\ref{aff13}}
\and A.~Pezzotta\orcid{0000-0003-0726-2268}\inst{\ref{aff17}}
\and M.~P\"ontinen\orcid{0000-0001-5442-2530}\inst{\ref{aff78}}
\and V.~Scottez\orcid{0009-0008-3864-940X}\inst{\ref{aff88},\ref{aff118}}
\and M.~Sereno\orcid{0000-0003-0302-0325}\inst{\ref{aff13},\ref{aff28}}
\and M.~Tenti\orcid{0000-0002-4254-5901}\inst{\ref{aff28}}
\and M.~Viel\orcid{0000-0002-2642-5707}\inst{\ref{aff7},\ref{aff5},\ref{aff26},\ref{aff6},\ref{aff119}}
\and M.~Wiesmann\orcid{0009-0000-8199-5860}\inst{\ref{aff69}}
\and Y.~Akrami\orcid{0000-0002-2407-7956}\inst{\ref{aff120},\ref{aff121}}
\and I.~T.~Andika\orcid{0000-0001-6102-9526}\inst{\ref{aff122},\ref{aff123}}
\and M.~Archidiacono\orcid{0000-0003-4952-9012}\inst{\ref{aff67},\ref{aff68}}
\and F.~Atrio-Barandela\orcid{0000-0002-2130-2513}\inst{\ref{aff124}}
\and P.~Bergamini\orcid{0000-0003-1383-9414}\inst{\ref{aff67},\ref{aff13}}
\and D.~Bertacca\orcid{0000-0002-2490-7139}\inst{\ref{aff1},\ref{aff65},\ref{aff2}}
\and M.~Bethermin\orcid{0000-0002-3915-2015}\inst{\ref{aff125}}
\and A.~Blanchard\orcid{0000-0001-8555-9003}\inst{\ref{aff98}}
\and L.~Blot\orcid{0000-0002-9622-7167}\inst{\ref{aff126},\ref{aff80}}
\and M.~Bonici\orcid{0000-0002-8430-126X}\inst{\ref{aff20},\ref{aff9}}
\and S.~Borgani\orcid{0000-0001-6151-6439}\inst{\ref{aff4},\ref{aff7},\ref{aff5},\ref{aff6},\ref{aff119}}
\and M.~L.~Brown\orcid{0000-0002-0370-8077}\inst{\ref{aff49}}
\and A.~Calabro\orcid{0000-0003-2536-1614}\inst{\ref{aff37}}
\and B.~Camacho~Quevedo\orcid{0000-0002-8789-4232}\inst{\ref{aff7},\ref{aff26},\ref{aff5}}
\and F.~Caro\inst{\ref{aff37}}
\and C.~S.~Carvalho\inst{\ref{aff104}}
\and T.~Castro\orcid{0000-0002-6292-3228}\inst{\ref{aff5},\ref{aff6},\ref{aff7},\ref{aff119}}
\and F.~Cogato\orcid{0000-0003-4632-6113}\inst{\ref{aff12},\ref{aff13}}
\and S.~Conseil\orcid{0000-0002-3657-4191}\inst{\ref{aff52}}
\and A.~R.~Cooray\orcid{0000-0002-3892-0190}\inst{\ref{aff127}}
\and O.~Cucciati\orcid{0000-0002-9336-7551}\inst{\ref{aff13}}
\and S.~Davini\orcid{0000-0003-3269-1718}\inst{\ref{aff14}}
\and G.~Desprez\orcid{0000-0001-8325-1742}\inst{\ref{aff128}}
\and A.~D\'iaz-S\'anchez\orcid{0000-0003-0748-4768}\inst{\ref{aff129}}
\and J.~J.~Diaz\orcid{0000-0003-2101-1078}\inst{\ref{aff47}}
\and S.~Di~Domizio\orcid{0000-0003-2863-5895}\inst{\ref{aff29},\ref{aff14}}
\and J.~M.~Diego\orcid{0000-0001-9065-3926}\inst{\ref{aff130}}
\and M.~Y.~Elkhashab\orcid{0000-0001-9306-2603}\inst{\ref{aff5},\ref{aff6},\ref{aff4},\ref{aff7}}
\and A.~Enia\orcid{0000-0002-0200-2857}\inst{\ref{aff13}}
\and Y.~Fang\orcid{0000-0002-0334-6950}\inst{\ref{aff66}}
\and A.~G.~Ferrari\orcid{0009-0005-5266-4110}\inst{\ref{aff28}}
\and A.~Finoguenov\orcid{0000-0002-4606-5403}\inst{\ref{aff78}}
\and A.~Fontana\orcid{0000-0003-3820-2823}\inst{\ref{aff37}}
\and A.~Franco\orcid{0000-0002-4761-366X}\inst{\ref{aff131},\ref{aff132},\ref{aff133}}
\and K.~Ganga\orcid{0000-0001-8159-8208}\inst{\ref{aff86}}
\and J.~Garc\'ia-Bellido\orcid{0000-0002-9370-8360}\inst{\ref{aff120}}
\and T.~Gasparetto\orcid{0000-0002-7913-4866}\inst{\ref{aff37}}
\and E.~Gaztanaga\orcid{0000-0001-9632-0815}\inst{\ref{aff43},\ref{aff44},\ref{aff134}}
\and F.~Giacomini\orcid{0000-0002-3129-2814}\inst{\ref{aff28}}
\and F.~Gianotti\orcid{0000-0003-4666-119X}\inst{\ref{aff13}}
\and G.~Gozaliasl\orcid{0000-0002-0236-919X}\inst{\ref{aff135},\ref{aff78}}
\and M.~Guidi\orcid{0000-0001-9408-1101}\inst{\ref{aff27},\ref{aff13}}
\and C.~M.~Gutierrez\orcid{0000-0001-7854-783X}\inst{\ref{aff136}}
\and A.~Hall\orcid{0000-0002-3139-8651}\inst{\ref{aff48}}
\and S.~Hemmati\orcid{0000-0003-2226-5395}\inst{\ref{aff24}}
\and H.~Hildebrandt\orcid{0000-0002-9814-3338}\inst{\ref{aff137}}
\and J.~Hjorth\orcid{0000-0002-4571-2306}\inst{\ref{aff93}}
\and S.~Joudaki\orcid{0000-0001-8820-673X}\inst{\ref{aff39}}
\and J.~J.~E.~Kajava\orcid{0000-0002-3010-8333}\inst{\ref{aff138},\ref{aff139}}
\and Y.~Kang\orcid{0009-0000-8588-7250}\inst{\ref{aff59}}
\and V.~Kansal\orcid{0000-0002-4008-6078}\inst{\ref{aff140},\ref{aff141}}
\and D.~Karagiannis\orcid{0000-0002-4927-0816}\inst{\ref{aff110},\ref{aff142}}
\and K.~Kiiveri\inst{\ref{aff76}}
\and J.~Kim\orcid{0000-0003-2776-2761}\inst{\ref{aff115}}
\and C.~C.~Kirkpatrick\inst{\ref{aff76}}
\and S.~Kruk\orcid{0000-0001-8010-8879}\inst{\ref{aff51}}
\and V.~Le~Brun\orcid{0000-0002-5027-1939}\inst{\ref{aff16}}
\and J.~Le~Graet\orcid{0000-0001-6523-7971}\inst{\ref{aff8}}
\and L.~Legrand\orcid{0000-0003-0610-5252}\inst{\ref{aff143},\ref{aff144}}
\and M.~Lembo\orcid{0000-0002-5271-5070}\inst{\ref{aff18},\ref{aff110},\ref{aff111}}
\and F.~Lepori\orcid{0009-0000-5061-7138}\inst{\ref{aff145}}
\and G.~Leroy\orcid{0009-0004-2523-4425}\inst{\ref{aff146},\ref{aff85}}
\and G.~F.~Lesci\orcid{0000-0002-4607-2830}\inst{\ref{aff12},\ref{aff13}}
\and J.~Lesgourgues\orcid{0000-0001-7627-353X}\inst{\ref{aff41}}
\and T.~I.~Liaudat\orcid{0000-0002-9104-314X}\inst{\ref{aff147}}
\and A.~Loureiro\orcid{0000-0002-4371-0876}\inst{\ref{aff148},\ref{aff149}}
\and J.~Macias-Perez\orcid{0000-0002-5385-2763}\inst{\ref{aff150}}
\and M.~Magliocchetti\orcid{0000-0001-9158-4838}\inst{\ref{aff61}}
\and C.~Mancini\orcid{0000-0002-4297-0561}\inst{\ref{aff9}}
\and F.~Mannucci\orcid{0000-0002-4803-2381}\inst{\ref{aff151}}
\and C.~J.~A.~P.~Martins\orcid{0000-0002-4886-9261}\inst{\ref{aff152},\ref{aff153}}
\and L.~Maurin\orcid{0000-0002-8406-0857}\inst{\ref{aff60}}
\and C.~J.~R.~McPartland\orcid{0000-0003-0639-025X}\inst{\ref{aff72},\ref{aff107}}
\and M.~Miluzio\inst{\ref{aff51},\ref{aff154}}
\and C.~Moretti\orcid{0000-0003-3314-8936}\inst{\ref{aff5},\ref{aff7},\ref{aff6},\ref{aff26}}
\and G.~Morgante\inst{\ref{aff13}}
\and S.~Nadathur\orcid{0000-0001-9070-3102}\inst{\ref{aff134}}
\and K.~Naidoo\orcid{0000-0002-9182-1802}\inst{\ref{aff134},\ref{aff75}}
\and P.~Natoli\orcid{0000-0003-0126-9100}\inst{\ref{aff110},\ref{aff111}}
\and A.~Navarro-Alsina\orcid{0000-0002-3173-2592}\inst{\ref{aff84}}
\and S.~Nesseris\orcid{0000-0002-0567-0324}\inst{\ref{aff120}}
\and D.~Paoletti\orcid{0000-0003-4761-6147}\inst{\ref{aff13},\ref{aff109}}
\and K.~Paterson\orcid{0000-0001-8340-3486}\inst{\ref{aff73}}
\and L.~Patrizii\inst{\ref{aff28}}
\and A.~Pisani\orcid{0000-0002-6146-4437}\inst{\ref{aff8}}
\and D.~Potter\orcid{0000-0002-0757-5195}\inst{\ref{aff145}}
\and S.~Quai\orcid{0000-0002-0449-8163}\inst{\ref{aff12},\ref{aff13}}
\and M.~Radovich\orcid{0000-0002-3585-866X}\inst{\ref{aff65}}
\and G.~Rodighiero\orcid{0000-0002-9415-2296}\inst{\ref{aff1},\ref{aff65}}
\and S.~Sacquegna\orcid{0000-0002-8433-6630}\inst{\ref{aff155}}
\and M.~Sahl\'en\orcid{0000-0003-0973-4804}\inst{\ref{aff156}}
\and D.~B.~Sanders\orcid{0000-0002-1233-9998}\inst{\ref{aff157}}
\and E.~Sarpa\orcid{0000-0002-1256-655X}\inst{\ref{aff26},\ref{aff119},\ref{aff6}}
\and A.~Schneider\orcid{0000-0001-7055-8104}\inst{\ref{aff145}}
\and M.~Schultheis\inst{\ref{aff114}}
\and D.~Sciotti\orcid{0009-0008-4519-2620}\inst{\ref{aff37},\ref{aff38}}
\and E.~Sellentin\inst{\ref{aff158},\ref{aff36}}
\and L.~C.~Smith\orcid{0000-0002-3259-2771}\inst{\ref{aff159}}
\and J.~G.~Sorce\orcid{0000-0002-2307-2432}\inst{\ref{aff160},\ref{aff60}}
\and K.~Tanidis\orcid{0000-0001-9843-5130}\inst{\ref{aff115}}
\and C.~Tao\orcid{0000-0001-7961-8177}\inst{\ref{aff8}}
\and G.~Testera\inst{\ref{aff14}}
\and R.~Teyssier\orcid{0000-0001-7689-0933}\inst{\ref{aff161}}
\and S.~Tosi\orcid{0000-0002-7275-9193}\inst{\ref{aff29},\ref{aff14},\ref{aff17}}
\and M.~Tucci\inst{\ref{aff59}}
\and D.~Vergani\orcid{0000-0003-0898-2216}\inst{\ref{aff13}}
\and G.~Verza\orcid{0000-0002-1886-8348}\inst{\ref{aff162}}
\and P.~Vielzeuf\orcid{0000-0003-2035-9339}\inst{\ref{aff8}}
\and N.~A.~Walton\orcid{0000-0003-3983-8778}\inst{\ref{aff159}}}
										   
\institute{Dipartimento di Fisica e Astronomia "G. Galilei", Universit\`a di Padova, Via Marzolo 8, 35131 Padova, Italy\label{aff1}
\and
INFN-Padova, Via Marzolo 8, 35131 Padova, Italy\label{aff2}
\and
Laboratoire Univers et Th\'eorie, Observatoire de Paris, Universit\'e PSL, Universit\'e Paris Cit\'e, CNRS, 92190 Meudon, France\label{aff3}
\and
Dipartimento di Fisica - Sezione di Astronomia, Universit\`a di Trieste, Via Tiepolo 11, 34131 Trieste, Italy\label{aff4}
\and
INAF-Osservatorio Astronomico di Trieste, Via G. B. Tiepolo 11, 34143 Trieste, Italy\label{aff5}
\and
INFN, Sezione di Trieste, Via Valerio 2, 34127 Trieste TS, Italy\label{aff6}
\and
IFPU, Institute for Fundamental Physics of the Universe, via Beirut 2, 34151 Trieste, Italy\label{aff7}
\and
Aix-Marseille Universit\'e, CNRS/IN2P3, CPPM, Marseille, France\label{aff8}
\and
INAF-IASF Milano, Via Alfonso Corti 12, 20133 Milano, Italy\label{aff9}
\and
Jet Propulsion Laboratory, California Institute of Technology, 4800 Oak Grove Drive, Pasadena, CA, 91109, USA\label{aff10}
\and
Kavli Institute for the Physics and Mathematics of the Universe (WPI), University of Tokyo, Kashiwa, Chiba 277-8583, Japan\label{aff11}
\and
Dipartimento di Fisica e Astronomia "Augusto Righi" - Alma Mater Studiorum Universit\`a di Bologna, via Piero Gobetti 93/2, 40129 Bologna, Italy\label{aff12}
\and
INAF-Osservatorio di Astrofisica e Scienza dello Spazio di Bologna, Via Piero Gobetti 93/3, 40129 Bologna, Italy\label{aff13}
\and
INFN-Sezione di Genova, Via Dodecaneso 33, 16146, Genova, Italy\label{aff14}
\and
California Institute of Technology, 1200 E California Blvd, Pasadena, CA 91125, USA\label{aff15}
\and
Aix-Marseille Universit\'e, CNRS, CNES, LAM, Marseille, France\label{aff16}
\and
INAF-Osservatorio Astronomico di Brera, Via Brera 28, 20122 Milano, Italy\label{aff17}
\and
Institut d'Astrophysique de Paris, UMR 7095, CNRS, and Sorbonne Universit\'e, 98 bis boulevard Arago, 75014 Paris, France\label{aff18}
\and
Ohio University, Physics \& Astronomy Department,1 Ohio University, Athens, OH 45701, USA\label{aff19}
\and
Waterloo Centre for Astrophysics, University of Waterloo, Waterloo, Ontario N2L 3G1, Canada\label{aff20}
\and
Department of Physics and Astronomy, University of Waterloo, Waterloo, Ontario N2L 3G1, Canada\label{aff21}
\and
Perimeter Institute for Theoretical Physics, Waterloo, Ontario N2L 2Y5, Canada\label{aff22}
\and
Minnesota Institute for Astrophysics, University of Minnesota, 116 Church St SE, Minneapolis, MN 55455, USA\label{aff23}
\and
Caltech/IPAC, 1200 E. California Blvd., Pasadena, CA 91125, USA\label{aff24}
\and
Universit\'e Paris-Saclay, Universit\'e Paris Cit\'e, CEA, CNRS, AIM, 91191, Gif-sur-Yvette, France\label{aff25}
\and
SISSA, International School for Advanced Studies, Via Bonomea 265, 34136 Trieste TS, Italy\label{aff26}
\and
Dipartimento di Fisica e Astronomia, Universit\`a di Bologna, Via Gobetti 93/2, 40129 Bologna, Italy\label{aff27}
\and
INFN-Sezione di Bologna, Viale Berti Pichat 6/2, 40127 Bologna, Italy\label{aff28}
\and
Dipartimento di Fisica, Universit\`a di Genova, Via Dodecaneso 33, 16146, Genova, Italy\label{aff29}
\and
Department of Physics "E. Pancini", University Federico II, Via Cinthia 6, 80126, Napoli, Italy\label{aff30}
\and
INAF-Osservatorio Astronomico di Capodimonte, Via Moiariello 16, 80131 Napoli, Italy\label{aff31}
\and
Dipartimento di Fisica, Universit\`a degli Studi di Torino, Via P. Giuria 1, 10125 Torino, Italy\label{aff32}
\and
INFN-Sezione di Torino, Via P. Giuria 1, 10125 Torino, Italy\label{aff33}
\and
INAF-Osservatorio Astrofisico di Torino, Via Osservatorio 20, 10025 Pino Torinese (TO), Italy\label{aff34}
\and
European Space Agency/ESTEC, Keplerlaan 1, 2201 AZ Noordwijk, The Netherlands\label{aff35}
\and
Leiden Observatory, Leiden University, Einsteinweg 55, 2333 CC Leiden, The Netherlands\label{aff36}
\and
INAF-Osservatorio Astronomico di Roma, Via Frascati 33, 00078 Monteporzio Catone, Italy\label{aff37}
\and
INFN-Sezione di Roma, Piazzale Aldo Moro, 2 - c/o Dipartimento di Fisica, Edificio G. Marconi, 00185 Roma, Italy\label{aff38}
\and
Centro de Investigaciones Energ\'eticas, Medioambientales y Tecnol\'ogicas (CIEMAT), Avenida Complutense 40, 28040 Madrid, Spain\label{aff39}
\and
Port d'Informaci\'{o} Cient\'{i}fica, Campus UAB, C. Albareda s/n, 08193 Bellaterra (Barcelona), Spain\label{aff40}
\and
Institute for Theoretical Particle Physics and Cosmology (TTK), RWTH Aachen University, 52056 Aachen, Germany\label{aff41}
\and
Deutsches Zentrum f\"ur Luft- und Raumfahrt e. V. (DLR), Linder H\"ohe, 51147 K\"oln, Germany\label{aff42}
\and
Institute of Space Sciences (ICE, CSIC), Campus UAB, Carrer de Can Magrans, s/n, 08193 Barcelona, Spain\label{aff43}
\and
Institut d'Estudis Espacials de Catalunya (IEEC),  Edifici RDIT, Campus UPC, 08860 Castelldefels, Barcelona, Spain\label{aff44}
\and
INFN section of Naples, Via Cinthia 6, 80126, Napoli, Italy\label{aff45}
\and
Dipartimento di Fisica e Astronomia "Augusto Righi" - Alma Mater Studiorum Universit\`a di Bologna, Viale Berti Pichat 6/2, 40127 Bologna, Italy\label{aff46}
\and
Instituto de Astrof\'{\i}sica de Canarias, E-38205 La Laguna, Tenerife, Spain\label{aff47}
\and
Institute for Astronomy, University of Edinburgh, Royal Observatory, Blackford Hill, Edinburgh EH9 3HJ, UK\label{aff48}
\and
Jodrell Bank Centre for Astrophysics, Department of Physics and Astronomy, University of Manchester, Oxford Road, Manchester M13 9PL, UK\label{aff49}
\and
European Space Agency/ESRIN, Largo Galileo Galilei 1, 00044 Frascati, Roma, Italy\label{aff50}
\and
ESAC/ESA, Camino Bajo del Castillo, s/n., Urb. Villafranca del Castillo, 28692 Villanueva de la Ca\~nada, Madrid, Spain\label{aff51}
\and
Universit\'e Claude Bernard Lyon 1, CNRS/IN2P3, IP2I Lyon, UMR 5822, Villeurbanne, F-69100, France\label{aff52}
\and
Institut de Ci\`{e}ncies del Cosmos (ICCUB), Universitat de Barcelona (IEEC-UB), Mart\'{i} i Franqu\`{e}s 1, 08028 Barcelona, Spain\label{aff53}
\and
Instituci\'o Catalana de Recerca i Estudis Avan\c{c}ats (ICREA), Passeig de Llu\'{\i}s Companys 23, 08010 Barcelona, Spain\label{aff54}
\and
Institut de Ciencies de l'Espai (IEEC-CSIC), Campus UAB, Carrer de Can Magrans, s/n Cerdanyola del Vall\'es, 08193 Barcelona, Spain\label{aff55}
\and
UCB Lyon 1, CNRS/IN2P3, IUF, IP2I Lyon, 4 rue Enrico Fermi, 69622 Villeurbanne, France\label{aff56}
\and
Departamento de F\'isica, Faculdade de Ci\^encias, Universidade de Lisboa, Edif\'icio C8, Campo Grande, PT1749-016 Lisboa, Portugal\label{aff57}
\and
Instituto de Astrof\'isica e Ci\^encias do Espa\c{c}o, Faculdade de Ci\^encias, Universidade de Lisboa, Campo Grande, 1749-016 Lisboa, Portugal\label{aff58}
\and
Department of Astronomy, University of Geneva, ch. d'Ecogia 16, 1290 Versoix, Switzerland\label{aff59}
\and
Universit\'e Paris-Saclay, CNRS, Institut d'astrophysique spatiale, 91405, Orsay, France\label{aff60}
\and
INAF-Istituto di Astrofisica e Planetologia Spaziali, via del Fosso del Cavaliere, 100, 00100 Roma, Italy\label{aff61}
\and
School of Physics, HH Wills Physics Laboratory, University of Bristol, Tyndall Avenue, Bristol, BS8 1TL, UK\label{aff62}
\and
University Observatory, LMU Faculty of Physics, Scheinerstr.~1, 81679 Munich, Germany\label{aff63}
\and
Max Planck Institute for Extraterrestrial Physics, Giessenbachstr. 1, 85748 Garching, Germany\label{aff64}
\and
INAF-Osservatorio Astronomico di Padova, Via dell'Osservatorio 5, 35122 Padova, Italy\label{aff65}
\and
Universit\"ats-Sternwarte M\"unchen, Fakult\"at f\"ur Physik, Ludwig-Maximilians-Universit\"at M\"unchen, Scheinerstr.~1, 81679 M\"unchen, Germany\label{aff66}
\and
Dipartimento di Fisica "Aldo Pontremoli", Universit\`a degli Studi di Milano, Via Celoria 16, 20133 Milano, Italy\label{aff67}
\and
INFN-Sezione di Milano, Via Celoria 16, 20133 Milano, Italy\label{aff68}
\and
Institute of Theoretical Astrophysics, University of Oslo, P.O. Box 1029 Blindern, 0315 Oslo, Norway\label{aff69}
\and
Felix Hormuth Engineering, Goethestr. 17, 69181 Leimen, Germany\label{aff70}
\and
Technical University of Denmark, Elektrovej 327, 2800 Kgs. Lyngby, Denmark\label{aff71}
\and
Cosmic Dawn Center (DAWN), Denmark\label{aff72}
\and
Max-Planck-Institut f\"ur Astronomie, K\"onigstuhl 17, 69117 Heidelberg, Germany\label{aff73}
\and
NASA Goddard Space Flight Center, Greenbelt, MD 20771, USA\label{aff74}
\and
Department of Physics and Astronomy, University College London, Gower Street, London WC1E 6BT, UK\label{aff75}
\and
Department of Physics and Helsinki Institute of Physics, Gustaf H\"allstr\"omin katu 2, University of Helsinki, 00014 Helsinki, Finland\label{aff76}
\and
Universit\'e de Gen\`eve, D\'epartement de Physique Th\'eorique and Centre for Astroparticle Physics, 24 quai Ernest-Ansermet, CH-1211 Gen\`eve 4, Switzerland\label{aff77}
\and
Department of Physics, P.O. Box 64, University of Helsinki, 00014 Helsinki, Finland\label{aff78}
\and
Helsinki Institute of Physics, Gustaf H{\"a}llstr{\"o}min katu 2, University of Helsinki, 00014 Helsinki, Finland\label{aff79}
\and
Laboratoire d'etude de l'Univers et des phenomenes eXtremes, Observatoire de Paris, Universit\'e PSL, Sorbonne Universit\'e, CNRS, 92190 Meudon, France\label{aff80}
\and
SKAO, Jodrell Bank, Lower Withington, Macclesfield SK11 9FT, UK\label{aff81}
\and
Centre de Calcul de l'IN2P3/CNRS, 21 avenue Pierre de Coubertin 69627 Villeurbanne Cedex, France\label{aff82}
\and
University of Applied Sciences and Arts of Northwestern Switzerland, School of Computer Science, 5210 Windisch, Switzerland\label{aff83}
\and
Universit\"at Bonn, Argelander-Institut f\"ur Astronomie, Auf dem H\"ugel 71, 53121 Bonn, Germany\label{aff84}
\and
Department of Physics, Institute for Computational Cosmology, Durham University, South Road, Durham, DH1 3LE, UK\label{aff85}
\and
Universit\'e Paris Cit\'e, CNRS, Astroparticule et Cosmologie, 75013 Paris, France\label{aff86}
\and
CNRS-UCB International Research Laboratory, Centre Pierre Bin\'etruy, IRL2007, CPB-IN2P3, Berkeley, USA\label{aff87}
\and
Institut d'Astrophysique de Paris, 98bis Boulevard Arago, 75014, Paris, France\label{aff88}
\and
Institute of Physics, Laboratory of Astrophysics, Ecole Polytechnique F\'ed\'erale de Lausanne (EPFL), Observatoire de Sauverny, 1290 Versoix, Switzerland\label{aff89}
\and
Telespazio UK S.L. for European Space Agency (ESA), Camino bajo del Castillo, s/n, Urbanizacion Villafranca del Castillo, Villanueva de la Ca\~nada, 28692 Madrid, Spain\label{aff90}
\and
Institut de F\'{i}sica d'Altes Energies (IFAE), The Barcelona Institute of Science and Technology, Campus UAB, 08193 Bellaterra (Barcelona), Spain\label{aff91}
\and
School of Mathematics and Physics, University of Surrey, Guildford, Surrey, GU2 7XH, UK\label{aff92}
\and
DARK, Niels Bohr Institute, University of Copenhagen, Jagtvej 155, 2200 Copenhagen, Denmark\label{aff93}
\and
Space Science Data Center, Italian Space Agency, via del Politecnico snc, 00133 Roma, Italy\label{aff94}
\and
Centre National d'Etudes Spatiales -- Centre spatial de Toulouse, 18 avenue Edouard Belin, 31401 Toulouse Cedex 9, France\label{aff95}
\and
Institute of Space Science, Str. Atomistilor, nr. 409 M\u{a}gurele, Ilfov, 077125, Romania\label{aff96}
\and
Institut f\"ur Theoretische Physik, University of Heidelberg, Philosophenweg 16, 69120 Heidelberg, Germany\label{aff97}
\and
Institut de Recherche en Astrophysique et Plan\'etologie (IRAP), Universit\'e de Toulouse, CNRS, UPS, CNES, 14 Av. Edouard Belin, 31400 Toulouse, France\label{aff98}
\and
Universit\'e St Joseph; Faculty of Sciences, Beirut, Lebanon\label{aff99}
\and
Departamento de F\'isica, FCFM, Universidad de Chile, Blanco Encalada 2008, Santiago, Chile\label{aff100}
\and
Universit\"at Innsbruck, Institut f\"ur Astro- und Teilchenphysik, Technikerstr. 25/8, 6020 Innsbruck, Austria\label{aff101}
\and
Satlantis, University Science Park, Sede Bld 48940, Leioa-Bilbao, Spain\label{aff102}
\and
Infrared Processing and Analysis Center, California Institute of Technology, Pasadena, CA 91125, USA\label{aff103}
\and
Instituto de Astrof\'isica e Ci\^encias do Espa\c{c}o, Faculdade de Ci\^encias, Universidade de Lisboa, Tapada da Ajuda, 1349-018 Lisboa, Portugal\label{aff104}
\and
Mullard Space Science Laboratory, University College London, Holmbury St Mary, Dorking, Surrey RH5 6NT, UK\label{aff105}
\and
Cosmic Dawn Center (DAWN)\label{aff106}
\and
Niels Bohr Institute, University of Copenhagen, Jagtvej 128, 2200 Copenhagen, Denmark\label{aff107}
\and
Universidad Polit\'ecnica de Cartagena, Departamento de Electr\'onica y Tecnolog\'ia de Computadoras,  Plaza del Hospital 1, 30202 Cartagena, Spain\label{aff108}
\and
INFN-Bologna, Via Irnerio 46, 40126 Bologna, Italy\label{aff109}
\and
Dipartimento di Fisica e Scienze della Terra, Universit\`a degli Studi di Ferrara, Via Giuseppe Saragat 1, 44122 Ferrara, Italy\label{aff110}
\and
Istituto Nazionale di Fisica Nucleare, Sezione di Ferrara, Via Giuseppe Saragat 1, 44122 Ferrara, Italy\label{aff111}
\and
INAF, Istituto di Radioastronomia, Via Piero Gobetti 101, 40129 Bologna, Italy\label{aff112}
\and
Astronomical Observatory of the Autonomous Region of the Aosta Valley (OAVdA), Loc. Lignan 39, I-11020, Nus (Aosta Valley), Italy\label{aff113}
\and
Universit\'e C\^{o}te d'Azur, Observatoire de la C\^{o}te d'Azur, CNRS, Laboratoire Lagrange, Bd de l'Observatoire, CS 34229, 06304 Nice cedex 4, France\label{aff114}
\and
Department of Physics, Oxford University, Keble Road, Oxford OX1 3RH, UK\label{aff115}
\and
Dipartimento di Fisica, Sapienza Universit\`a di Roma, Piazzale Aldo Moro 2, 00185 Roma, Italy\label{aff116}
\and
Aurora Technology for European Space Agency (ESA), Camino bajo del Castillo, s/n, Urbanizacion Villafranca del Castillo, Villanueva de la Ca\~nada, 28692 Madrid, Spain\label{aff117}
\and
ICL, Junia, Universit\'e Catholique de Lille, LITL, 59000 Lille, France\label{aff118}
\and
ICSC - Centro Nazionale di Ricerca in High Performance Computing, Big Data e Quantum Computing, Via Magnanelli 2, Bologna, Italy\label{aff119}
\and
Instituto de F\'isica Te\'orica UAM-CSIC, Campus de Cantoblanco, 28049 Madrid, Spain\label{aff120}
\and
CERCA/ISO, Department of Physics, Case Western Reserve University, 10900 Euclid Avenue, Cleveland, OH 44106, USA\label{aff121}
\and
Technical University of Munich, TUM School of Natural Sciences, Physics Department, James-Franck-Str.~1, 85748 Garching, Germany\label{aff122}
\and
Max-Planck-Institut f\"ur Astrophysik, Karl-Schwarzschild-Str.~1, 85748 Garching, Germany\label{aff123}
\and
Departamento de F{\'\i}sica Fundamental. Universidad de Salamanca. Plaza de la Merced s/n. 37008 Salamanca, Spain\label{aff124}
\and
Universit\'e de Strasbourg, CNRS, Observatoire astronomique de Strasbourg, UMR 7550, 67000 Strasbourg, France\label{aff125}
\and
Center for Data-Driven Discovery, Kavli IPMU (WPI), UTIAS, The University of Tokyo, Kashiwa, Chiba 277-8583, Japan\label{aff126}
\and
Department of Physics \& Astronomy, University of California Irvine, Irvine CA 92697, USA\label{aff127}
\and
Kapteyn Astronomical Institute, University of Groningen, PO Box 800, 9700 AV Groningen, The Netherlands\label{aff128}
\and
Departamento F\'isica Aplicada, Universidad Polit\'ecnica de Cartagena, Campus Muralla del Mar, 30202 Cartagena, Murcia, Spain\label{aff129}
\and
Instituto de F\'isica de Cantabria, Edificio Juan Jord\'a, Avenida de los Castros, 39005 Santander, Spain\label{aff130}
\and
INFN, Sezione di Lecce, Via per Arnesano, CP-193, 73100, Lecce, Italy\label{aff131}
\and
Department of Mathematics and Physics E. De Giorgi, University of Salento, Via per Arnesano, CP-I93, 73100, Lecce, Italy\label{aff132}
\and
INAF-Sezione di Lecce, c/o Dipartimento Matematica e Fisica, Via per Arnesano, 73100, Lecce, Italy\label{aff133}
\and
Institute of Cosmology and Gravitation, University of Portsmouth, Portsmouth PO1 3FX, UK\label{aff134}
\and
Department of Computer Science, Aalto University, PO Box 15400, Espoo, FI-00 076, Finland\label{aff135}
\and
 Instituto de Astrof\'{\i}sica de Canarias, E-38205 La Laguna; Universidad de La Laguna, Dpto. Astrof\'\i sica, E-38206 La Laguna, Tenerife, Spain\label{aff136}
\and
Ruhr University Bochum, Faculty of Physics and Astronomy, Astronomical Institute (AIRUB), German Centre for Cosmological Lensing (GCCL), 44780 Bochum, Germany\label{aff137}
\and
Department of Physics and Astronomy, Vesilinnantie 5, University of Turku, 20014 Turku, Finland\label{aff138}
\and
Serco for European Space Agency (ESA), Camino bajo del Castillo, s/n, Urbanizacion Villafranca del Castillo, Villanueva de la Ca\~nada, 28692 Madrid, Spain\label{aff139}
\and
ARC Centre of Excellence for Dark Matter Particle Physics, Melbourne, Australia\label{aff140}
\and
Centre for Astrophysics \& Supercomputing, Swinburne University of Technology,  Hawthorn, Victoria 3122, Australia\label{aff141}
\and
Department of Physics and Astronomy, University of the Western Cape, Bellville, Cape Town, 7535, South Africa\label{aff142}
\and
DAMTP, Centre for Mathematical Sciences, Wilberforce Road, Cambridge CB3 0WA, UK\label{aff143}
\and
Kavli Institute for Cosmology Cambridge, Madingley Road, Cambridge, CB3 0HA, UK\label{aff144}
\and
Department of Astrophysics, University of Zurich, Winterthurerstrasse 190, 8057 Zurich, Switzerland\label{aff145}
\and
Department of Physics, Centre for Extragalactic Astronomy, Durham University, South Road, Durham, DH1 3LE, UK\label{aff146}
\and
IRFU, CEA, Universit\'e Paris-Saclay 91191 Gif-sur-Yvette Cedex, France\label{aff147}
\and
Oskar Klein Centre for Cosmoparticle Physics, Department of Physics, Stockholm University, Stockholm, SE-106 91, Sweden\label{aff148}
\and
Astrophysics Group, Blackett Laboratory, Imperial College London, London SW7 2AZ, UK\label{aff149}
\and
Univ. Grenoble Alpes, CNRS, Grenoble INP, LPSC-IN2P3, 53, Avenue des Martyrs, 38000, Grenoble, France\label{aff150}
\and
INAF-Osservatorio Astrofisico di Arcetri, Largo E. Fermi 5, 50125, Firenze, Italy\label{aff151}
\and
Centro de Astrof\'{\i}sica da Universidade do Porto, Rua das Estrelas, 4150-762 Porto, Portugal\label{aff152}
\and
Instituto de Astrof\'isica e Ci\^encias do Espa\c{c}o, Universidade do Porto, CAUP, Rua das Estrelas, PT4150-762 Porto, Portugal\label{aff153}
\and
HE Space for European Space Agency (ESA), Camino bajo del Castillo, s/n, Urbanizacion Villafranca del Castillo, Villanueva de la Ca\~nada, 28692 Madrid, Spain\label{aff154}
\and
INAF - Osservatorio Astronomico d'Abruzzo, Via Maggini, 64100, Teramo, Italy\label{aff155}
\and
Theoretical astrophysics, Department of Physics and Astronomy, Uppsala University, Box 516, 751 37 Uppsala, Sweden\label{aff156}
\and
Institute for Astronomy, University of Hawaii, 2680 Woodlawn Drive, Honolulu, HI 96822, USA\label{aff157}
\and
Mathematical Institute, University of Leiden, Einsteinweg 55, 2333 CA Leiden, The Netherlands\label{aff158}
\and
Institute of Astronomy, University of Cambridge, Madingley Road, Cambridge CB3 0HA, UK\label{aff159}
\and
Univ. Lille, CNRS, Centrale Lille, UMR 9189 CRIStAL, 59000 Lille, France\label{aff160}
\and
Department of Astrophysical Sciences, Peyton Hall, Princeton University, Princeton, NJ 08544, USA\label{aff161}
\and
Center for Computational Astrophysics, Flatiron Institute, 162 5th Avenue, 10010, New York, NY, USA\label{aff162}}    
\title{\Euclid preparation}
\subtitle{First investigation of the impact of cross-contamination on spectroscopic redshift measurements with pixel-level simulations}

\abstract{We present a study on simulated data focused on understanding the performance of the spectroscopic redshift measurements with the \gls*{nisp} instrument on \Euclid. Simulations include scenarios with different levels of cross-contamination arising from overlapping spectra of nearby sources, which represents one of the main drawbacks of slitless spectroscopy. We present a new analysis based on pixel-level simulations of the NISP images, with the data processed using the \Euclid spectroscopic pipeline. We first consider an idealised case with non-overlapping spectra to assess the accuracy and reliability of the redshift measurement as a function of the flux of the \halpha emission line and galaxy size. We then introduce more realistic contamination scenarios, distinguishing between two contributions: contamination from \halpha emitters, which are the \Euclid targets for cosmological analyses, and contamination from all other galaxies. In the second case, we analyse the impact of cross-contamination with an increasing number of contaminants, from the brighter to the fainter galaxies. Given that our results show no clear evidence that sources fainter than magnitude 20 degrade redshift measurements, we conservatively restrict our analysis to galaxies with magnitudes up to 24. In particular, we provide a preliminary estimate that contamination from galaxies within the same redshift range as the target sample contributes to about 4\% of the total degradation due to cross-contamination from all galaxies.}


\titlerunning{Impact of cross-contamination on spectroscopic redshifts estimation}

\maketitle

\section{\label{sc:Intro}Introduction}

The observation of light distributions at large scales provides crucial insights into the evolution of the Universe. Galaxy surveys perform cosmological investigations by using galaxies as tracers to map the large-scale structure. \Euclid (\citealp{Laureijs11, EuclidSkyOverview}) will contribute to this effort, providing one of the largest three-dimensional maps of galaxies in terms of sky coverage and redshift range. The \Euclid spacecraft is equipped with two scientific instruments: the Visible Imager (VIS; \citealp{EuclidSkyVIS}) and the Near-Infrared Spectrometer and Photometer (NISP; \citealp{EuclidSkyNISP}).
They are expected to provide measurements of the shapes and magnitudes for billions of objects, along with accurate spectroscopic redshifts for tens of millions of them. Statistical analysis of these measurements through weak lensing and galaxy clustering will provide us with precious cosmological information.

\Euclid spectroscopic observations are performed with a slitless instrument, meaning that the light from all objects in the \gls*{fov} is dispersed without a pre-selection of the target sources. This enables wide-field observations and allows for a simple instrument design, making it ideal for large space-based surveys. The slitless approach has been adopted in previous space missions, such as the \HST \citep{Walsh2010, Freudling2008, Kummel2006-ACS, Kuntschner2010, Bagley2020}, and it is planned to be used in future instruments such as the \textit{Nancy Grace Roman} Space Telescope \citep{Wang2022}, enabling wide-field spectroscopy at near-infrared wavelengths, which are challenging with ground-based observations. However, estimating redshifts through slitless spectroscopy suffers from three main drawbacks \citep{Kummel2009, Outini2020}: \begin{inparaenum}[(a)] \item a high background light; \item a degradation of the effective spectral resolution with the object size, called `self-contamination'; \item the overlap of spectra from nearby objects, known as `cross-contamination'. \end{inparaenum} All these effects have an impact on the redshift measurement and must be taken into account in the selection of the data sample for the cosmological analysis. Background light and self-contamination lead to a reduction of the \gls*{snr}, which can be characterised with simplified simulations (Euclid Collaboration: Granett et al., in prep.). Modelling the effect of cross-contamination is more complex as it depends on several factors: the two-dimensional density of sources in the sky, the intensity and shape of the contaminants, and the dispersion angles of light. Therefore, comprehensive pixel-level simulations are required to accurately assess these systematic effects.  

In \Euclid, spectroscopic redshifts are primarily determined by detecting the \halpha emission line, which is typically the brightest feature in the optical rest-frame spectra of emission-line galaxies. Within the \gls*{nisp} wavelength coverage (\qty{900}{nm}--\qty{2000}{nm}), \halpha emissions can be observed in galaxies with redshifts in the range $0.9 \leq z \leq 1.8$ \citep{EuclidSkyNISP}. The main survey, referred to as the \gls*{ews}, is designed to achieve a \gls*{snr} of 3.5 for an extended source of \halpha flux $\qty{2e-16}{erg\,  s^{-1} cm^{-2}}$ and radius $r \leq \ang{;;0.25}$. We introduce a dimensionless rescaled flux variable, denoted ${\cal F}$, defined as 
\begin{equation}
{\fcal} := \frac{F}{10^{-16}\,\mathrm{erg\,s^{-1}\,cm^{-2}}}
= \frac{F}{10^{-19}\,\mathrm{W\,m^{-2}}}\,,
\end{equation} where $F$ is the line flux. This normalization provides a convenient scaling such that typical flux values are of order unity and is used throughout this paper.

Incorrect redshift measurements due to noise fluctuations in the spectra, misclassification of the \halpha line with other emission lines, and residual artefacts from the spectroscopic processing lead to interlopers in the catalogue used for cosmological analysis. The impact of these interlopers must be carefully assessed, as it can distort the three-dimensional distribution of galaxies with consequences on the correlation function and power spectrum (\citealp{EP-Risso},  Euclid Collaboration: Lee et al., in prep.). The fraction of interlopers depends on the \gls*{snr} of the spectra in the sample: a higher \gls*{snr} spectrum makes both the \halpha and other lines easier to detect, leading to a lower probability of redshift error due to both random fluctuations and line misidentifications. The \gls*{snr} is determined by several factors such as the flux of each emission line, the galaxy size, the Milky Way extinction, and the background light. Cross-contamination strongly impacts the number of interlopers by reducing the \gls*{snr} of the emission lines, giving residuals from the decontamination process, and lowering the fraction of usable pixels in each spectrum. \bigskip

In this paper, we present a study based on spectroscopic pixel-level simulations aimed at assessing the performance of the redshift measurement for the \Euclid \halpha target sample. Although pixel-level simulations are time-consuming and involve complex configurations for each processing step, they are a key tool to test spectral reconstruction pipelines and investigate effects that are difficult to reproduce with other tools. Previous works, such as \cite{Gabarra-EP31} and \cite{EP-Lusso}, use this type of simulations to assess the performance of the \Euclid spectroscopy in the absence of cross-contamination. We extend the existing simulation framework to enable spectral decontamination within the \Euclid pipeline and redshift measurements.

In particular, we focus on measuring the statistical uncertainty in the redshift measurement and we evaluate the success rate as a function of the line flux and galaxy size. In addition, we perform a new analysis to assess the impact of cross-contamination as a function of the brightness and redshift of the contaminant sources, which is a critical aspect of slitless spectroscopy and has not yet been studied with simulations. We also give a first estimate of the effect of contaminant sources that lie at the same redshift as the target \halpha emitters. In fact, while cross-contamination from foreground sources can be mitigated (upon verification they are not correlated with the \halpha galaxies) by constructing a suitable random catalogue that takes into account this effect, contamination from sources that lie at the same redshift creates a complicated selection bias that requires specific mitigation strategies.

In this analysis, we make exclusive use of the official \Euclid pipeline for the simulation of the images \citep{EP-Serrano}, spectral extraction \citep{Q1-TP006}, and redshift measurement \citep{Q1-TP007}. Given the computational demands of a full end-to-end pipeline of both photometric and spectroscopic observations, we adopt some simplifications. We limit the analysis to the spectroscopic channel, using idealised photometric cutouts and assuming perfect photometric and morphological measurements. In addition, we ignore the zeroth and second orders of spectra. External algorithms such as \texttt{grizli} \citep{Brammer2021} are tested on \Euclid simulations, and results are presented in Euclid Collaboration: McCarthy et al. (in prep.).

Although the pipeline continues to improve and simulations rely on some simplifications, this analysis provides important insights into specific systematic effects. In particular, it allows these effects to be examined separately, unlike in real observations where systematics are entangled with each other. It should be noted that the results presented here do not represent a definitive assessment of redshift performance, given the assumptions and idealisations regarding the simulation of the sources, instrumental modelling, and data processing. Rather, this study is a qualitative investigation to support the development of the mitigation strategies, which can be refined with further simulations and observed data.

\bigskip
This paper is structured as follows. Section \ref{sc:2} presents the sample used for the cosmological analysis and the input catalogue used for the pixel-level simulations. Section \ref{sc:3} describes the simulations and reconstruction pipelines: the selection of the observations, the spectroscopic simulator, the photometric bypass, the extraction of the spectra, and the redshift measurement. A first set of simulations without cross-contamination is presented in Sect. \ref{sc:4}. This analysis provides a measurement of the redshift success rate as a function of both the intensity of the \halpha emission line and the size of the galaxies. Simulations with cross-contamination at different magnitudes are described in Sect. \ref{sc:5}. The paper concludes with an overall discussion of results in Sect. \ref{sc:6}. All magnitude values are reported using the AB magnitude system \citep{Schirmer-EP18}. 

\section{\label{sc:2} Galaxy distribution and contamination in \Euclid}

\subsection{\halpha and contaminant samples}\label{sec:2.1}
In the \gls*{ews}, the \gls*{nisp} instrument is expected to achieve a limiting magnitude of 24.0 in the three photometric bands \YE, \JE, and \HE, with a \gls*{snr} of 5 for point-like sources \citep{Scaramella-EP1}. In the spectroscopic channel, only sources with continuum in the  \HE band brighter than magnitude 19.5 are expected to reach a \gls*{snr} greater than 3.5 \citep{Gabarra-EP31}. Based on the photometric detection limit, we restrict our analysis to galaxies with $\HE < 24$. We further divide the selected galaxy distribution into two samples:
\begin{itemize}
    \item \halpha sample: galaxies with redshift $z \in [0.9, 1.8]$ and flux of the $\mathrm{H\alpha +[\ion{N}{ii}]}$ complex
    $\fhalphacal \geq 1$.
    This flux threshold is a factor of two below the nominal detection limit in the \gls*{nisp} spectroscopic channel. Although only a small fraction of sources with flux 
    $\fhalphacal \in [1, \, 2]$
    is expected to be correctly measured, their inclusion in our simulations provides a more complete characterisation of systematics for the fainter sources.  

    It is important to highlight that, to properly characterise the \Euclid capability to detect the \halpha emitters, in this work we will often select, for detection, only objects with flux 
    $\fhalphacal \geq 2$.
    
    \bigskip
    \item Contaminant sample: all galaxies that are not in the  \halpha sample. 
\end{itemize}

In this paper, we distinguish between simulated quantities and quantities estimated through the redshift estimation pipeline using the subscripts `true' and `meas', respectively. Furthermore, when the context is not ambiguous, we omit the `true' subscript for simplicity.\bigskip

The measured \halpha sample used for cosmological inference is both impure and incomplete \citep{EP-Monaco2}. Sources in the contaminant sample contribute to this impurity and incompleteness. The purity of the observed \halpha sample crucially depends on the relative abundance of true contaminants compared to true \halpha emitters. For instance, even assuming that all the redshifts of the \halphat sample are correctly identified, if (as expected) \halpha emitters are approximately $2\%$ of the total sample, the wrong identification of just $1\%$ of true contaminants as \halpha galaxies would reduce the purity to $67\%$.

Contaminant sources degrade the quality of the measured \halpha sample due to cross-contamination, contributing to its incompleteness. The level of degradation depends on the number and intensity of contaminated pixels. Therefore, extended and bright sources are more likely to introduce significant contamination in nearby spectra than compact or faint ones. Cross-contamination is modelled and subtracted for each spectrum during data processing; the \gls*{snr} in the affected pixels is reduced, and residuals from the subtraction may remain. Both continuum and emission lines from contaminating sources contribute to such degradation: contamination from the continuum affects a larger fraction of pixels with a lower intensity, whereas contamination from emission lines is more localised but stronger. Since the expected fraction of emission line galaxies detectable with \gls*{nisp} is only $\sim 2\%$ of the total number of galaxies, and most of them are \halpha emitters, the remaining contaminants contribute to the degradation of the signal through their continuum. Even if only a small fraction of contaminants has a detectable continuum, their large number could lead to a significant increase in the noise. Therefore, in this work, we focus specifically on the effect of cross-contamination arising from the continuum of contaminants as a function of magnitude (and as a consequence of their number).

Optimising the purity and completeness of the cosmological sample requires a meticulous analysis of the selection criteria. In this work, we focus on quantifying the impact on the redshift measurements of cross-contamination due to spectral continua. Selection of the cosmological sample and assessment of the purity and completeness are addressed in \cite{Cagliari24} and Euclid Collaboration: McCarthy et al. (in prep.).

\subsection{Effect of cross-contamination on the cosmological analysis}
\begin{figure}
    \centering
    \includegraphics[width=\linewidth]{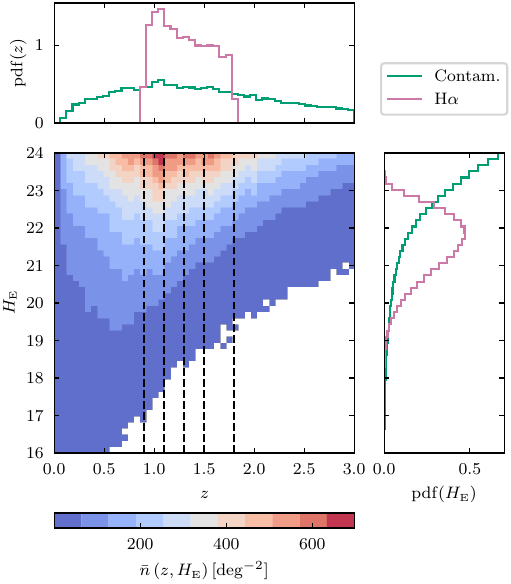}
    \caption{Expected average number density of galaxies per square degree detectable with \gls*{nisp} as a function of \HE magnitude and true redshift. The distribution is computed from a $\sim \qty{36}{deg^2}$ area described in Sect.~\ref{sc:3.1}. The black dashed lines indicate the values of the four redshift bins used in the galaxy clustering analysis. The one-dimensional histograms in the top and right panels display the distributions of the true redshift and \HE\ magnitude, respectively, with the area of each histogram normalised to one.  The \halpha sample is shown in pink, while the contaminant sample is shown in green.}
    \label{fig:2_mag_vs_z}
\end{figure}

To mitigate systematic effects arising from cross-contamination, it is crucial to understand which sources significantly affect the redshift measurement. Figure \ref{fig:2_mag_vs_z} shows the expected average number density of galaxies per square degree $\bar{n}\, (z, \HE)$ as a function of redshift and magnitude. Bright sources are predominantly located at low redshift, while a larger population of faint sources is located at around the \Euclid target redshift range. Nevertheless, due to the faint continua of sources at high redshift, the dominant contribution to cross-contamination is expected to arise from bright and low-redshift galaxies.

Particular attention must be given to the redshift distribution of galaxies that significantly affect redshift measurements due to cross-contamination. When the spectrum of an \halpha galaxy is contaminated by the signal of a nearby object on the sky, the probability of correctly measuring its redshift is reduced. This effect results in an observed decrease in the number density of galaxies used for the cosmological analysis. If the redshift of the contaminants is $z \leq 0.9$ or $z \geq 1.8$, this is located in very distant regions of the Universe with respect to the one sampled by target galaxies, so (at least within the $\Lambda$CDM model) we can assume that the target and contaminant galaxies are statistically independent. In this case, provided we identified the contaminants, the observed density fluctuations caused by contamination can be mitigated by introducing a spatial modulation in the visibility mask, which is represented by a random catalogue subject to the same selection effects as the data catalogue \citep{EP-Monaco2}. It should be noted that, while the observed density can be corrected, the loss of signal leads to an increase in the noise. However, this mitigation strategy works as long as the contamination is statistically independent of the density field we are measuring. If contaminants lie at redshift $0.9 < z < 1.8$, this assumption is broken. This leads to a density-dependent selection effect, where denser regions may have fewer correct redshift measurements. In this scenario, the resulting underdensity cannot be accounted for by the visibility mask, and a dedicated mitigation strategy must be developed to address this systematic effect.

Spectra of \halpha galaxies may also overlap with each other on the focal plane. In this case, given their low number density and faint continuum, their cross-contamination is expected to be negligible. Within our working assumptions, we will confirm this expectation.

\subsection{\label{sc:input_catalog}Input catalogue for simulations}
To simplify the analysis, we simulate only galaxies. We assume that it is possible to perfectly distinguish between stars and galaxies. Thus, since we consider stars as independent of the cosmological signal, the impact of their cross-contamination on the redshift measurement can be properly treated during the mitigation of systematics, with the same approach described for galaxies at redshift $z\le 0.9$ or $z\ge 1.8$.

We assume the Flagship galaxy mock described in \cite{EuclidSkyFlagship} as the true galaxy distribution. The Flagship simulation was created to reproduce a catalogue of galaxies expected to be observed with \Euclid. The catalogue is based on an $N$-body dark matter simulation with 4 trillion particles, from which 16 billion dark matter haloes were identified. Dark matter haloes were then populated with galaxies through halo occupation distribution models and abundance matching calibrated on observed properties such as luminosity, spectral energy distribution, shapes, and emission-line fluxes. The final output is a catalogue of 3.4 billion galaxies covering one octant of the sky up to redshift $z=3$ and with magnitude $\HE < 26$. The output catalogue is distributed through the CosmoHub platform \citep{TALLADA2020100391, 2017ehep.confE.488C} 

Being the first study on the effect of cross-contamination, we adopt a simplified framework. We retain the clustering of \halpha emitters and randomise the angular positions of the contaminant sample to remove their intrinsic clustering. This should have a negligible impact when the redshift of the contaminant galaxies is $z\le 0.9$ or $z\ge 1.8$. However, this is not obviously the case for the contamination from galaxies lying at the same redshift as the \halpha sample, as they will likely show a similar clustering pattern. A detailed analysis of the impact of clustering properties on cross-contamination, including their dependence on the object type, is beyond the scope of this work and is left to future studies.

\section{\label{sc:3}Pixel-level simulations}
\subsection{\label{sc:3.1}Selection of the observations}
\begin{figure}
    \centering
    \includegraphics[width=.5\textwidth]{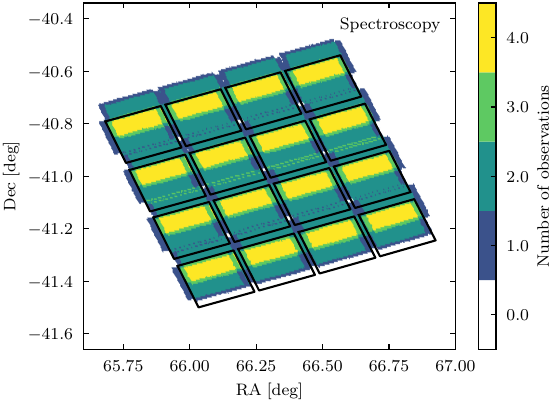}
    \caption{Spectroscopic coverage for a single observation. The black lines outline the \gls*{nisp} detectors projected in the sky for the first pointing. The colour bar reports the positions on the sky for which the spectrum is fully observed the relative number of times.}
    \label{fig:coverage}
\end{figure}

Each \gls*{ews} observation is realised with four dithered pointings \citep{Scaramella-EP1}. Due to the presence of gaps between the detectors of the \gls*{nisp} focal plane, not all sources are observed in all four pointings.  Only  $\sim 50\%$ of sources are expected to be observed four times or more. Coverage above four is reached for sources falling within the small overlap regions of contiguous pointings. This effect is even more significant in spectroscopy: since each spectrum extends over approximately 530 pixels, we estimate that about 50\% of spectra will be observed at least 3 times or more in all the wavelengths covered, while about 32\% will be covered four times or more. In this work, we do not simulate contiguous observations, further reducing the coverage at the edges of the focal plane. This choice allows us to study fields with different densities of \halpha galaxies while avoiding the additional complexity introduced by overlapping observations. Figure~\ref{fig:coverage} shows the spectroscopic coverage for one observation. 

\begin{figure}
    \centering
    \includegraphics[width=\linewidth]{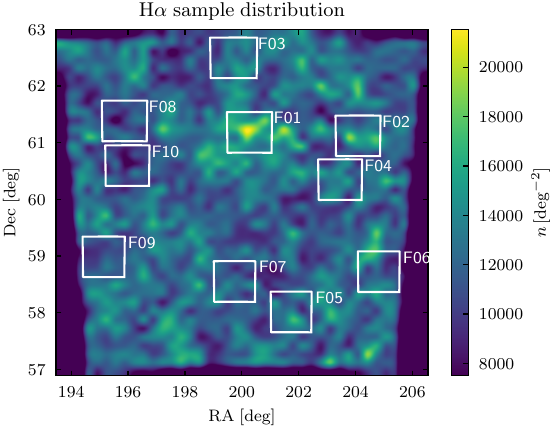}
    \caption{Two-dimensional spatial distribution of the \halpha sample. White boxes outline the projection of NISP focal plane for the 10 simulated observations.}
    \label{fig:halpha_spatial_distribution}
\end{figure}

\bigskip  
Considering the required efforts, we decided to simulate 10 field observations. We selected galaxies and observations to simulate from a $\qty{36}{deg^2}$ region of the Flagship galaxy mock, in order to exhibit a large variance in the surface number density of the targets. This choice maximises the range of local \halpha densities represented by the simulations, ensuring the presence of both underdense (voids) and overdense (clusters) regions (see Fig.~\ref{fig:halpha_spatial_distribution}). Contaminant galaxies, assigned random angular positions, are unaffected by this selection. Since, as shown below, our results are insensitive to the local \halpha density (see Fig.~\ref{fig:eff_target}) we conclude that there are no reasons why the chosen region on the Flagship should bias the results of our investigation. We uniformly sampled the two-dimensional spatial distribution of the \halpha galaxies, shown in  Figure~\ref{fig:halpha_spatial_distribution}, with a grid of $20\times20$ pixels of sky coordinates. From this grid, we selected 10 coordinates as the centre of the observations in order to cover the full range of galaxy density, from underdense to overdense, and avoiding overlaps. Additional and technical details are given in Appendix \ref{apx:1}. The resulting observations are shown in Fig. \ref{fig:halpha_spatial_distribution} and the corresponding number densities are reported in Table \ref{table:obs_coordinates}.

\subsection{Simulations and reconstruction with SPRING}
\begin{figure*}
\centering
\includegraphics[width=\linewidth]{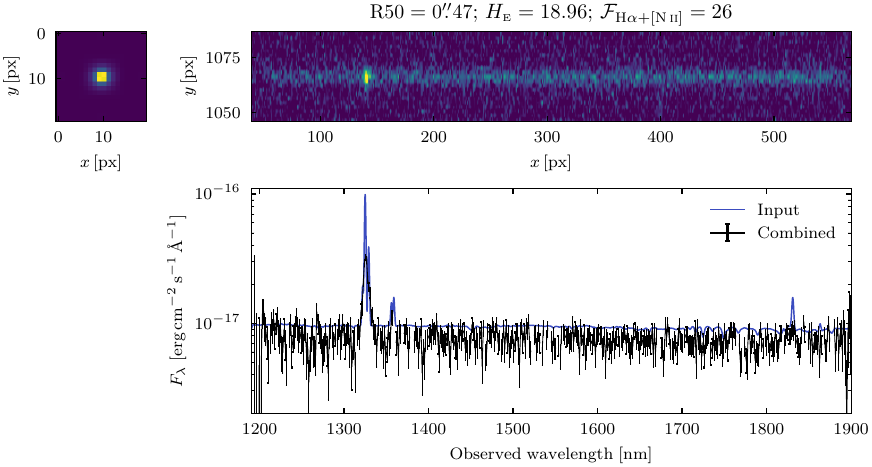}
\caption{Example of a simulated spectrum. \emph{Top left}: idealised photometric cutout. \emph{Top right}: 2D spectrum of a single pointing. \emph{Lower panel}: input and combined spectra as a function of the observed wavelengths. In this example, the continuum and the \halpha emission-line fluxes are particularly bright. 
}
\label{fig:spectrum_esample}
\end{figure*}

To evaluate the performance and possible systematic effects induced by the \Euclid data processing, it is necessary to execute the complete and official pipelines, from the simulation of the spectra up to the redshift measurement. The Spectroscopic Pipeline Runner and Input Generator (\texttt{SPRING}; \citealp{spring_abstract}) has been designed to meet this purpose. Giving the proper input (catalogues, instrument calibrations, survey strategy, configurations on the pipelines) it produces dispersed images using the \texttt{TIPS} simulator \citep{EuclidSkyOverview}, extracts 1D spectra using the \texttt{SIR} pipeline \citep{Q1-TP006}, and measures redshifts and spectroscopic features with the \texttt{SPE} pipeline \citep{Q1-TP007}. An example of a simulated and reconstructed spectrum with \texttt{SPRING} is shown in Fig. \ref{fig:spectrum_esample}. \texttt{SPRING} is an updated version of \texttt{SIR\_SpectroSim\_Runner} \citep{Paganin2022phd}, which has already been used in previous studies to evaluate the performance of the \Euclid spectroscopic channel \citep{Gabarra-EP31, EP-Lusso}. Compared to the previous code, \texttt{SPRING} introduces important features: \begin{inparaenum}[(a)] \item the integration with the \Euclid 
\glsentrylong{ial} (\glsentryshort{ial}; \citealt{Frailis2019}), which handles the management of computational tasks; \item the support of parallel subprocess execution through IAL, reducing the execution time by a factor 4 to 16, and enabling simulations over larger sky areas and number of sources;
\item the generation of simplified photometric cutouts to run the decontamination algorithm in \texttt{SIR}, which is one of the most challenging aspects of slitless spectroscopy;
\item the execution of a full end-to-end spectroscopic pipeline, including the redshift and spectral feature measurement with the integration of the \texttt{SPE} pipeline.
\end{inparaenum}

In this study, we use the following versions of the pipelines: \texttt{TIPS} 9.2.7, \texttt{SIR} 4.6.2, and \texttt{SPE} 1.1. It is important to highlight that the \Euclid pipeline was in development during the period of this work; nevertheless, the processing is similar to the one used for \Euclid Q1 data. Future releases are expected to provide improvements in the quality of the results. However, processing data with these pipelines will allow testing the capabilities and obtaining preliminary quantification of the performances and of possible systematic effects introduced by them.

\subsubsection{NISP-S simulations with \texttt{TIPS}}
\texttt{TIPS} \citep{Zoubian2014, EP-Serrano} is the \Euclid simulator for the \gls*{nisp} spectroscopic channel. Instrument parameters were calibrated during ground-test campaigns, allowing the modelling of detector properties in terms of dark current, readout noise, and quantum efficiency at the pixel level (\citealp[]{EU-Kubik}, \citealp[]{EP-Gillard}). These effects exhibit a spatial variation across the focal plane with an average value of $\qty{0.03}{\mathrm{e}^{-}\; s^{-1}\;\mathrm{px}^{-1}}$ of dark current, 95\% of quantum efficiency, and $\qty{13}{\mathrm{e}^{-}}$ of readout noise. Significant spatial variations are observed between different detectors, particularly in dark current and quantum efficiency, and are corrected using pixel-level calibration maps. In the simulated exposures, the acquisition of the signal follows the  Multi-ACCumulation (MACC) scheme used in-flight and presented in \cite{Kubik2016}.

The simulator accounts for different astrophysical effects, including cosmic ray hits and background light. Cosmic ray events are modelled according to the expected energy spectrum from \texttt{CREME96} \citep{Creme97, Creme2010, Creme2012}. Background light in \texttt{TIPS} simulations includes contributions from zodiacal and out-of-field stray light. The total background, estimated following  \cite{Scaramella-EP1}, is assumed to be uniform across the  \gls*{fov}, and it is determined exclusively by the coordinates of the centre of the pointing. In the sky region selected for the observations and described in Sect.~\ref{sc:3.1}, the average zodiacal light and out-of-field stray light are $\qty{1.3}{e^{-1} \, s^{-1} \,  px^{-1}}$ and $\qty{0.3}{e^{-1} \, s^{-1} \,  px^{-1}}$ respectively.

At the time of this study, accurate calibrations of the zeroth order spectra were not available. The zeroth order exhibits a characteristic double-peaked structure extending over approximately 10 pixels \citep{Q1-TP006}. Considering the lack of accurate calibration and our main interest on the continuum, we restricted the simulations and analysis to the first-order signal.

\subsubsection{Photometric bypass}
Photometric observations with the VIS \citep{EuclidSkyVIS} and NISP \citep{EuclidSkyNISP} instruments are processed independently through dedicated pipelines \citep{Q1-TP003, Q1-TP002}. This first processing produces fully calibrated exposures along with preliminary catalogues. The calibrated frames are then analysed by the \texttt{MER} processing function to provide image mosaics, together with photometric and morphological catalogues of the sources \citep{Q1-TP004}.

Simulating the complete data acquisition would require not only the simulation of VIS and NISP photometry, but also processing them with \texttt{VIS}, \texttt{NIR}, and \texttt{MER} pipelines. This approach has two issues: it is computationally demanding, and it requires knowledge of the details of all involved pipelines. To overcome these limitations, \texttt{SPRING} has been developed to avoid running the simulation and processing of the photometric channels. Rather than simulating these observations, \texttt{SPRING} creates idealised photometric cutouts directly from the morphological parameters listed in the Flagship galaxy mocks. Cutouts are created considering the photometric point spread function, 
without introducing any instrumental noise. Photometric cutouts are generated with the \texttt{GalSim} package \citep{Rowe2015}, at a spatial resolution of $\ang{;;0.3}$. In addition to the photometric cutouts, the \texttt{SIR} pipeline requires information regarding the morphological parameters and the fluxes. In this simplified framework, this information is not computed from the imaging, but it is taken from the Flagship galaxy mocks. Thus, photometric errors, such as those affecting the source detection, flux and size measurement, are not included. The lack of photometric simulations and processing is the only difference between the \texttt{SPRING} pipeline and the full \Euclid pipeline. Our analysis thus likely underestimates errors and overestimates the \gls*{snr}. Also for this reason, as mentioned in Sect~\Ref{sc:Intro}, our investigation will give optimistic results. Those are nevertheless precious to support the development of more refined future analyses.

\subsubsection{Extraction of 1D spectra}\label{sec:3-sir}
The dispersed images generated with \texttt{TIPS} are processed through the \texttt{SIR} pipeline, which performs several tasks: image pre-processing, spectral decontamination to correct for overlapping spectra, wavelength and flux calibrations, and the extraction of one-dimensional spectra \citep{Q1-TP006}. 

The spectral extraction uses an aperture whose size depends on the galaxy morphology, with a minimum size of 5 pixels. The aperture size is optimised to simultaneously minimise the flux loss and maximise the \gls*{snr}. Even for a point-like source, this configuration leads to a flux loss of $\sim 5\%$ with a 5-pixel aperture \citep{Paganin2022phd}. This flux loss increases with galaxy size as shown in \cite{Gabarra-EP31}. One of the planned improvements in the \texttt{SIR} pipeline is the implementation of the optimal profile-weighted extraction \citep{Robertson86, Horne98} to correct this loss. At the time of this work, this method is still under development and testing, and thus, flux losses are expected in the spectra. 

The one-dimensional calibrated spectra from individual exposures are combined using inverse-variance weighting, with outliers rejected with a pull-clipping \citep{Q1-TP006}. The final outputs of the \texttt{SIR} pipeline are one-dimensional combined spectra, with associated flux, variance, and mask flagging problematic pixels. The resulting spectra cover the wavelength range from $\qty{1190}{nm}$ to $\qty{1900}{nm}$ with a wavelength step of $\qty{1.34}{nm}$.

\subsubsection{Redshift measurement}
Measurements on the extracted \Euclid spectra are performed using the \texttt{SPE} processing function described in \cite{Q1-TP007}. Spectroscopic redshifts are measured
through a fit over a grid of spectral templates evaluated at different redshifts. For each model, a redshift probability density function ($z$PDF) is computed, and different solutions are ranked based on their probability. In this analysis, only the most probable solution for each galaxy, corresponding to the largest value of the parameter \texttt{spe\_z\_prob} introduced below, is considered.

An important feature of the computation of the spectroscopic redshift performed in \texttt{SPE} is the inclusion of a prior that favours solutions containing the \halpha line in the \gls*{nisp} wavelength range. With this choice, it is more probable to identify a single emission line as an \halpha, leading to a gain in completeness at the cost of a reduction in purity. Studying the impact of different priors is out of the scope of this paper. For the processing, we applied the previously mentioned prior on the \halpha emission line as defined for the \gls*{ews}. 

Given the large size of the \Euclid data sample, robust metrics are required to assess the reliability of the redshift measurement. The \texttt{SPE} pipeline provides the \texttt{spe\_z\_prob} parameter, which is the integral of the $z$PDF over a $\pm3\sigma$ interval around the main peak (where $\sigma$ is the width of the Gaussian fit to the peak). A spectrum with multiple emission lines will probably have a single peak in the $z$PDF with an integral close to 1, while a spectrum with no evident emission lines will show multiple peaks in the $z$PDF with a smaller value of the integral. Results presented in \cite{Q1-TP007} show that the parameter \texttt{spe\_z\_prob} must be close to unity to ensure robust redshift estimates, with solutions below 0.99 not reliable.

\section{\label{sc:4}Simulations on the grid for the \halpha sample}
\begin{figure}
    \centering
    \includegraphics[width=.8\linewidth]{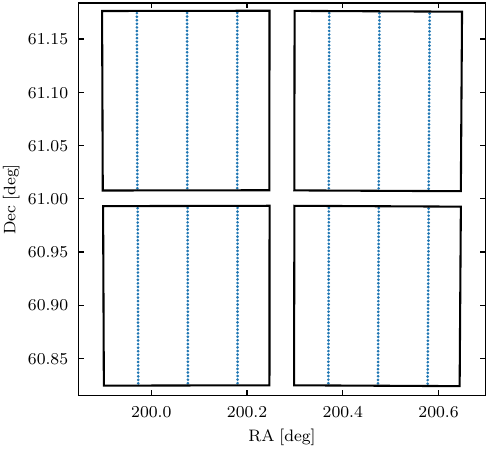}
    \caption{Spatial distribution of simulated sources arranged on the grid. Blue markers correspond to their position in the sky. In this configuration, spectra are approximately dispersed along the RA axis. The black outlines show the sky projection of 4 out of the 16 detectors of the NISP instrument.}
    \label{fig:fov-grid}
\end{figure}

We produced simplified simulations using the same framework as the complete analysis, but without cross-contamination. These simulations allow us to estimate redshift measurement accuracy and provide a benchmark for comparison with other analyses of the \Euclid spectroscopic performance. These simulations consist of 6 observations with a dithering step of only 5 pixels, designed to prevent spectra from falling in the gaps of the NISP instrument and to ensure that bad pixels do not affect the spectra multiple times in the same region. In this approach, all spectra are observed four times in their full wavelength range. Galaxies are arranged `on the grid' with $50 \times 3$ spectra in each detector to simultaneously maximise the number of simulated objects and avoid cross-contamination. The source coordinates and the projection of the focal plane on the sky are shown in Fig. \ref{fig:fov-grid}. 

\begin{figure}
    \centering
    \includegraphics[width=\linewidth]{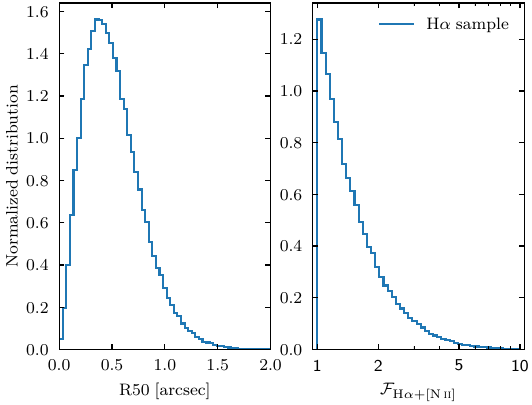}
    \caption{Properties of the \halpha sample. \emph{Left panel}: density distribution of R50, corresponding to the half-light radius as described in the text. \emph{Right panel}: flux of the \halpha complex. The areas of the histograms are normalised to one.}
    \label{fig:halpha_size_and_flux}
\end{figure}

Only galaxies from the \halphat sample were simulated. Spatial profiles and flux of the \halpha complex were simulated according to the Flagship galaxy mock with a bulge plus disc component. We report in Fig.~\ref{fig:halpha_size_and_flux} the distributions of half-light radius and flux of the \halpha complex. We describe the galaxy size with the parameter R50, which corresponds to the half-light radius of the disc component when present, or the half-light radius of the bulge component otherwise. A total of $\qty{13147}{}$ sources were simulated\footnote{This is slightly less than the number that we could have accommodated.}.

\subsection{\label{sec:z_accuracy}Redshift accuracy}
\begin{figure}[]
     \centering
     \includegraphics[width=\linewidth]{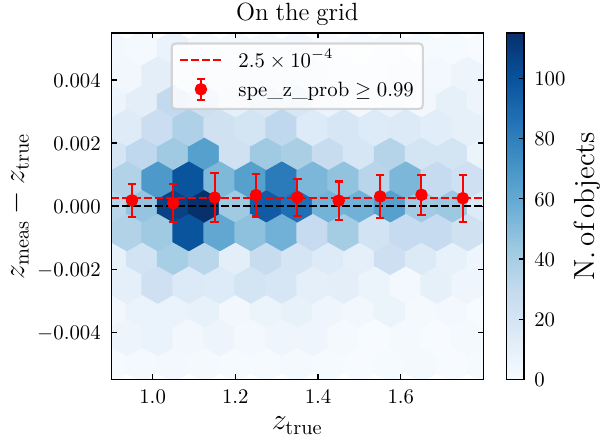}
    \caption{Distribution of $\delta z$ as a function of $z_\mathrm{true}$. The two-dimensional histogram shows the measurements for all simulated \halpha galaxies. Red markers and error bars represent the median and \gls*{mad} with a 0.1 redshift bin width. Median and \gls*{mad} are computed on a subsample that has $\texttt{spe\_z\_prob}\geq 0.99$.  The dashed black line corresponds to $\delta z=0$.}
    \label{fig:delta_z_grid_vs_z}
\end{figure}

In this section, we present the results on the accuracy of the redshift measurement, which are summarised in the plot in Fig. \ref{fig:delta_z_grid_vs_z}. The plot shows the difference $\delta z = z_{\mathrm{meas}} - z_{\mathrm{true}}$ between the measured and true redshift as a function of the true one. The two-dimensional histogram shows the distribution of all simulated \halpha emitters included in the \halpha sample, without any selection criteria. Red markers indicate the median $\delta z$ for reliable galaxies ($\texttt{spe\_z\_prob} \geq 0.99$), binned in intervals of $z_{\mathrm{true}}$ of 0.1 width. Error bars report the median absolute deviation (MAD) within each bin. We employ the median and not the mean because of the large number of outliers, as it will be explained later in this section. The plot shows a bias of $\qty{2.5e-4}{}$ in $\delta z$, which appears independent of the true redshift. The bias is likely due to a systematic shift to longer wavelengths of the \halpha+[\ion{N}{ii}] complex in cases of large [\ion{N}{ii}] fluxes. Additional details on the bias are discussed in Appendix \ref{apx:2}. 

\begin{figure*}[]
     \centering
     \includegraphics[width=\linewidth]{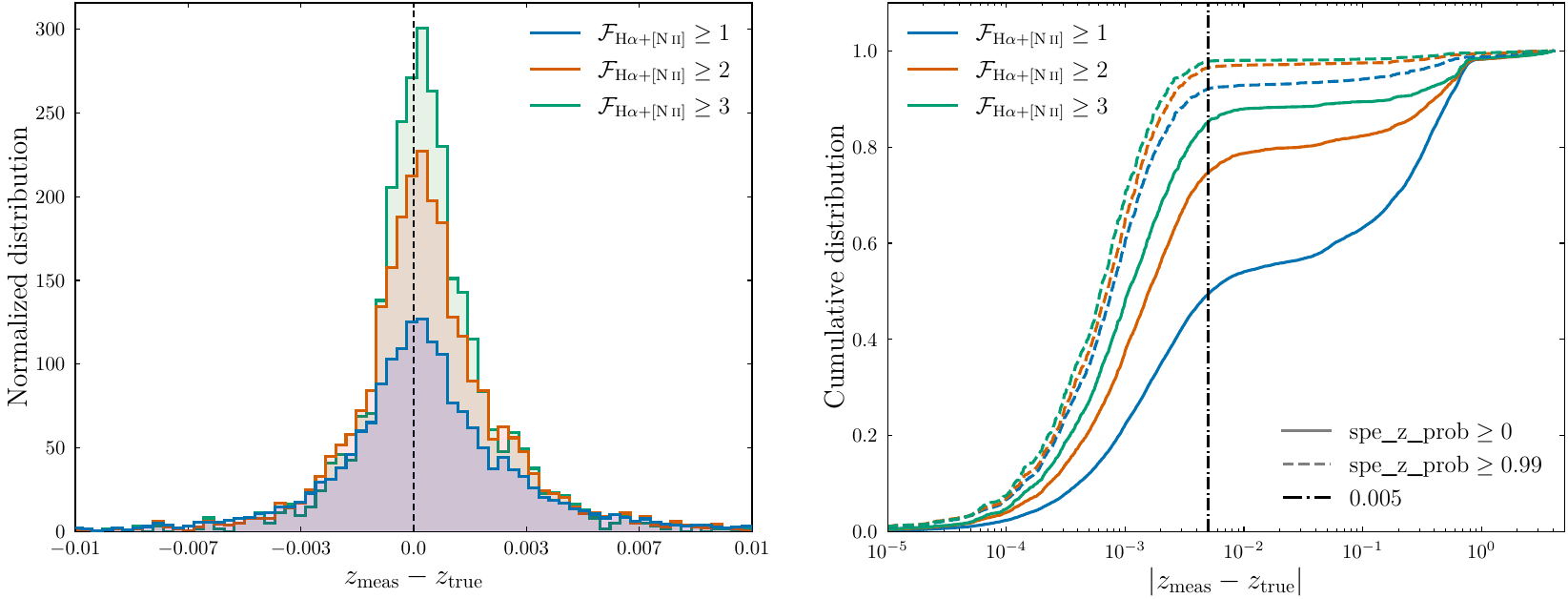}
    \caption{Redshift accuracy for galaxies on the grid. \emph{Left panel}: density distributions of $z_{\mathrm{meas}} - z_{\mathrm{true}}$ for three different samples: 
    $\fhalphacal \geq 1$
    in blue,  
    $\fhalphacal \geq 2$
    in orange, and 
    $\fhalphacal \geq 3$
    in green. No cut on \texttt{spe\_z\_prob} is applied in this case. The areas of the histograms are normalised to one.  \emph{Right panel}: cumulative distribution of $|z_{\mathrm{meas}} - z_{\mathrm{true}}|$ for the same flux limits. The solid lines represent measurements without a reliability constraint, while dashed lines display measurements with $\texttt{spe\_z\_prob} \geq 0.99$. The vertical black dash-dotted line exhibits the cut-off threshold between the core and tails of the samples at $|z_{\mathrm{meas}} - z_{\mathrm{true}}|=0.005$.}
    \label{fig:delta_z_grid}
\end{figure*}

Figure~\ref{fig:delta_z_grid} displays the $\delta z$ distribution for different samples. Samples are selected by applying different cuts on the lower limit of simulated flux. The left panel shows the distributions without any selection based on the redshift reliability, while the right panel presents the cumulative distributions of $\vert \delta z \vert$. All distributions exhibit two components: a `core', which corresponds to correctly measured redshifts, and extended tails caused by outliers originating from line misidentifications and noise fluctuations in the spectra. Distinguishing between these two components is not straightforward; nevertheless, it is crucial to define a threshold to distinguish between properly measured redshifts and wrong measurements. For instance, a simple $3\sigma$ cut is not meaningful in this context, as the central peak of the distribution is not well described by a Gaussian profile; tails also extend at high $\vert \delta z \vert$ values.  
To move forward, we consider the sample selected based on the \gls*{nisp} nominal \halpha flux limit 
($\fhalphacal \ge 2$)
and restrict the analysis to reliable measurements ($\texttt{spe\_z\_prob} \ge 0.99$). We find that 93\%, 95\%, and 97\% of the measurements have $\vert \delta z \vert$ smaller than 0.003, 0.004, and 0.005 respectively, while 2\% of the measurements have $\vert \delta z \vert\ge 0.2$. Based on these results and supported by the visual inspection of the distributions in Fig.~\ref{fig:delta_z_grid}, we adopt a threshold of $\vert \delta z \vert = 0.005$ to identify outliers. While the width of the distributions exhibits variations depending on the selected emission line flux, the cumulative distributions (see Fig.~\ref{fig:delta_z_grid}) indicate that small changes to the $\vert \delta z \vert =$ threshold (of about 0.001--0.002) do not significantly affect the resulting fraction of selected sources. Therefore, from here on, we consider accurate redshift measurement, those which satisfy $|z_{\mathrm{meas}} - z_{\mathrm{true}}| \leq 0.005$.

The number of sources in each sample and the fraction of measurements considered correct are reported in Table \ref{tab:delta_z_grid}. As already discussed, for the galaxies on the grid, 97\% of \halpha galaxies with flux 
$\fhalphacal \geq 2$
and $\texttt{spe\_z\_prob} \geq 0.99$ are accurately measured. This fraction drops to 75\% if no reliability criterion is applied. However, higher purity comes at a cost: the cut on \texttt{spe\_z\_prob} for galaxies with 
$\fhalphacal \geq 2$ 
is fulfilled only by 29\% of sources. As a result, 63\% of galaxies with accurately measured redshifts\footnote{The fraction of accurately measured galaxies that are excluded when applying the \texttt{spe\_z\_prob} reliability cut is computed as the difference between the number of accurate measurements without the cut and with the cut, divided by the number without the cut.} and 97\% of interlopers are excluded.

Apart from the small bias in redshift measurement for large [\ion{N}{ii}] fluxes in the simulations, the results obtained in this work are not in conflict with those presented in \cite{Q1-TP007}. Nevertheless, this bias is not relevant for the present investigation as it does not affect our results.

\begin{table}[h]
    \caption{Summary statistics on the redshift measurement. The table reports the number and the fraction of accurate redshift measurements for six subsamples, defined by different flux limits of the \halpha complex and thresholds on the   \texttt{spe\_z\_prob} reliability parameter.}
    \centering
    \begin{tabular}{|c|c|c|c|}
        \hline 
        \rule{0pt}{3.5ex}
        ${\cal F}_\mathrm{{H\alpha+[\ion{N}{ii}]}}$ & spe\_z\_prob\ & N. sources & \% with $|\delta z| \leq 0.005$ \rule[-1.5ex]{0pt}{0pt} \\ 
        \hline 
        $1$ & $0$ & 13126 & $49\%$ \\
        $2$ & $0$ & 4549 & $75\%$ \\
        $3$ & $0$ & 2042 & $85\%$ \\
        $1$ & $0.99$ & 1625 & $92\%$ \\
        $2$ & $0.99$ & 1312 & $97\%$ \\
        $3$ & $0.99$ & 863 & $98\%$ \\
        \hline 
    \end{tabular}\\[9pt]
    \label{tab:delta_z_grid}
\end{table}

\subsection{\label{sec:4.2}Redshift success rate} 
The fraction of correct redshift measurements depends on the flux of the \halpha emission line, as well as quality cuts as \gls*{snr}, \texttt{spe\_z\_prob}, galaxy size, and photometry \citep{Cagliari24}. We studied the simulation without cross-contamination to investigate the dependence of the redshift performance on both the \halpha flux and the size of the sources. We report the results in terms of Redshift Success Rate ($\mathcal{SR}$), defined as

\begin{equation}\label{eq:1}
    \mathcal{SR} = \frac{N\; \bigg(\mathrm{H\alpha_{true}}\; \&\; z_\text{accurate}  \; \&\; z_\text{reliable} \;\&\; \mathcal{S}e\ell
 \bigg)}{N \; \bigg(\mathrm{H\alpha_{true}} \; \&\; \mathcal{S}e\ell\bigg) }
\end{equation}

where:
\begin{itemize}
    \item $\mathrm{H\alpha_{true}}$: galaxies in the \halpha sample, i.e. those with $0.9 \leq z_\text{true} \leq 1.8$ and 
    ${\cal F}_{\mathrm{H\alpha} + [\mathrm{\ion{N}{ii}}],\, {\rm true}} \geq 2$,
    as defined in Sect. \ref{sec:2.1};
    \item $z_\text{accurate}$: measurements which satisfy $|z_{\mathrm{meas}} - z_{\mathrm{true}}| \leq \qty{5e-3}{}$;
    \item $z_\text{reliable}$: measurement considered reliable, i.e. $\texttt{spe\_z\_prob} \geq 0.99$;
    \item $\mathcal{S}e\ell$: additional selection criteria, such as the 
    number of exposures covering the full redshift range, the galaxy size and the \halpha flux. For instance, a selection $\mathcal{S}e\ell$ used in this paper includes 4 observations, $\ang{;;0} \leq \text{R50} < \ang{;;0.25}$ and an emission-line flux greater than  
    ${\cal F}_{\mathrm{H\alpha} + [\mathrm{\ion{N}{ii}}],\, {\rm true}} \geq 2$.
    Specific selection criteria will be defined in each case.
\end{itemize}

This success rate is strictly connected to the measured \halpha galaxy number density $n_0(\vec{x})$ per unit of comoving volume, where $\vec{x}$ is the comoving coordinate. Following \cite{EP-Monaco2}, $n_0$ can be expressed as the integral over the \halpha emission line flux $F$ of the local luminosity function  
$\phi_{\rm local}(F\vert \vec{x})$ -- i.e.~the number of galaxies per unit of $F$ and comoving volume -- and a completeness function estimated as $\mathcal{SR}(F, z)$ from Eq.~\eqref{eq:1} ,
\begin{equation}
    n_{\rm o}(\vec{x}) = \int_0^\infty  \mathcal{SR}(F, z)\, \phi_{\rm local}(F\vert \vec{x})\, {\rm d}F,
\end{equation}
where we assume the success rate does not depend on the comoving position $\vec{x}$ but just on the emission line flux and redshift.

\begin{figure*}
     \centering
    \includegraphics[width=\linewidth]{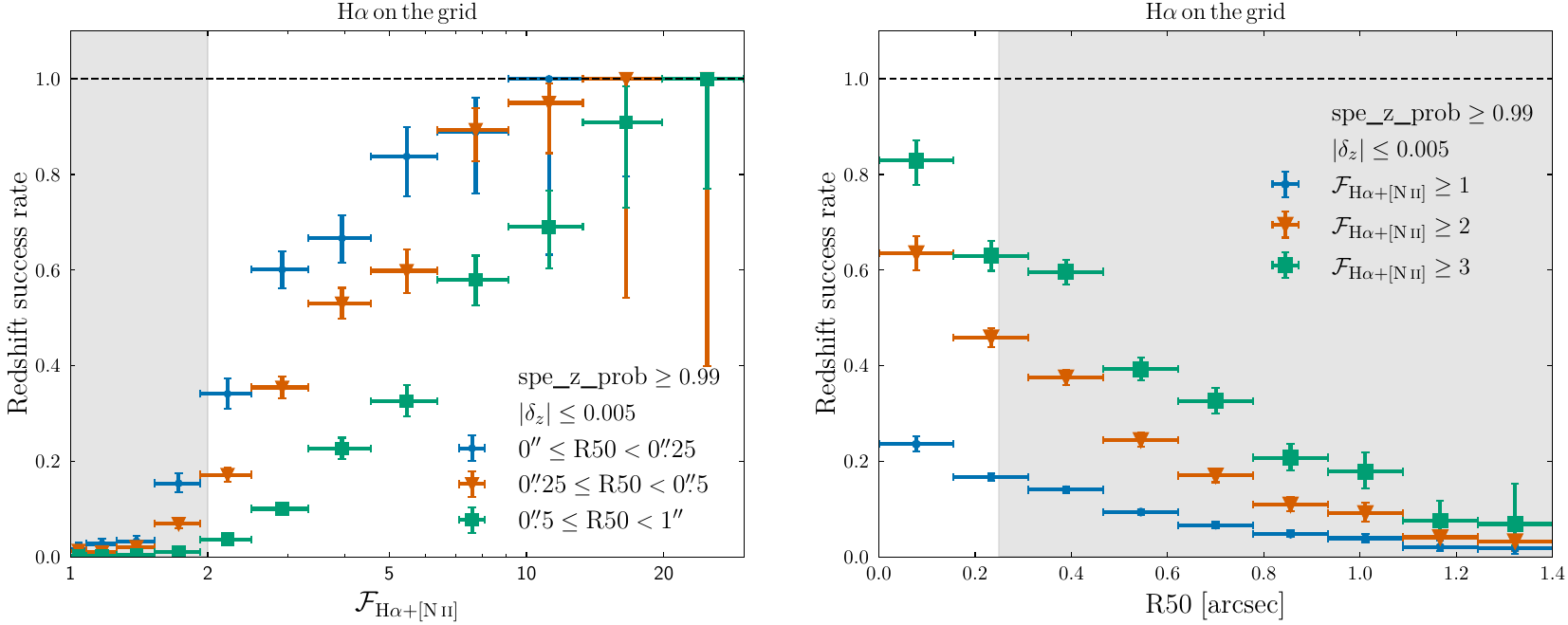}
   \caption{\emph{Left panel}: redshift success rate as a function of the input flux of the \halpha complex. Three different sub-samples are reported with different colours: $\ang{;;0} \leq \text{R50} < \ang{;;0.25}$ (blue), $\ang{;;0.25} \leq \text{R50} < \ang{;;0.5}$ (orange), and $\ang{;;0.5} \leq \text{R50} < \ang{;;1}$ (green). \emph{Right panel}: redshift success rate as a function of the size of the galaxies. Subsamples for 
   $\fhalphacal \geq 1$
   (blue), 
   $\fhalphacal \geq 2$
   (orange), and 
   $\fhalphacal \geq 3$
   (green) are shown. The shaded grey regions in the left and right panels show values of \halpha flux and half-light radius, for which \halpha emission lines are expected to be less detectable.}\label{fig:sr_vs_flux_and_size}
\end{figure*}

Results on the success rate are presented in Fig. \ref{fig:sr_vs_flux_and_size}. For galaxies on the grid, the selection $\mathcal{S}e\ell$ implicitly includes the requirement that each spectrum is observed four times within the full wavelength range. The error bars on the success rate are Clopper--Pearson intervals \citep{ClopperPearson1934} at the 68\% confidence level, while the horizontal error bars show the bin widths used in the computation. We adopted Clopper--Pearson intervals because it is a method for calculating binomial confidence intervals that guarantees coverage and is particularly useful when the success probability is near 0 or 1. It is worth noting that the error on the redshift measurement $\delta z$, quantified in Sect.~\ref{sec:z_accuracy}, is taken into account in the computation of the value of the success rate, but it does not propagate into its error bars. The success rate errors are mainly determined by the number of galaxies we consider; they indeed decrease with increasing sample size and approach zero in the limit of an infinite sample.

The left panel shows the redshift success rate as a function of the true flux $\fhalpha$ flux for three subsamples of different galaxy sizes. For 
$\fhalphacal < 2$
only a small fraction of redshifts are measured correctly and marked as reliable: for sources with $\mathrm{R50} < \ang{;;0.25}$, the success rate is 2\% for 
$\fhalphacal \in [1, \, 1.4]$,
and 10\% for 
$\fhalphacal \in [1.4, \,1.9]$. 
For strong \halpha emission lines, the success rate reaches 100\% between 
$\fhalphacal = 10$ and $\fhalphacal = 20$
depending on the size of the source.

The right panel reports the success rate as a function of the R50 parameter. In this case, samples are selected applying different cuts on the lower limit of the simulated  \halpha flux. As already suggested by the left plot, the galaxy size significantly impacts the redshift measurement. For galaxies with 
$\fhalphacal \geq 2$, 
the success rate goes from 55\% to 37\% when passing from $\text{R50} < \ang{;;0.25}$ to  $\ang{;;0.25} \leq \text{R50} < \ang{;;0.5}$. The shaded grey regions in the left and right panels show values of \halpha flux and half-light radius, respectively, for which \halpha emission lines are expected to be less detectable due to the instrumental design.

\section{\label{sc:5} Simulations of \halpha sample with contamination}
We simulated 10 observations selected from the Flagship galaxy mock as described in Sect.~\ref{sc:input_catalog}, to evaluate the effect of cross-contamination. We produced multiple simulations with an increasing number of sources for each observation. We started simulating only \halpha galaxies, and then added contaminant galaxies brighter than a certain magnitude; magnitude thresholds were set at \{19, 20, 21, 22, 23, 24\} mag, resulting in a total of 70 simulations.

\subsection{\label{sec:5.1} Contamination within \halpha galaxies}
In this section, we evaluate the impact of cross-contamination due to \halpha galaxies only.  On average, each spectrum of an \halpha emitter is contaminated by 4 to 7 other spectra per exposure, depending on the number density of the observation. Moreover, as shown in Fig. \ref{fig:2_mag_vs_z}, the average \HE magnitude of \halpha galaxies is about 21.5, which is relatively faint. As a consequence, contamination within the \halpha sample is not expected to significantly affect the redshift measurement.

\begin{figure*}
     \centering
    \includegraphics[width=\linewidth]{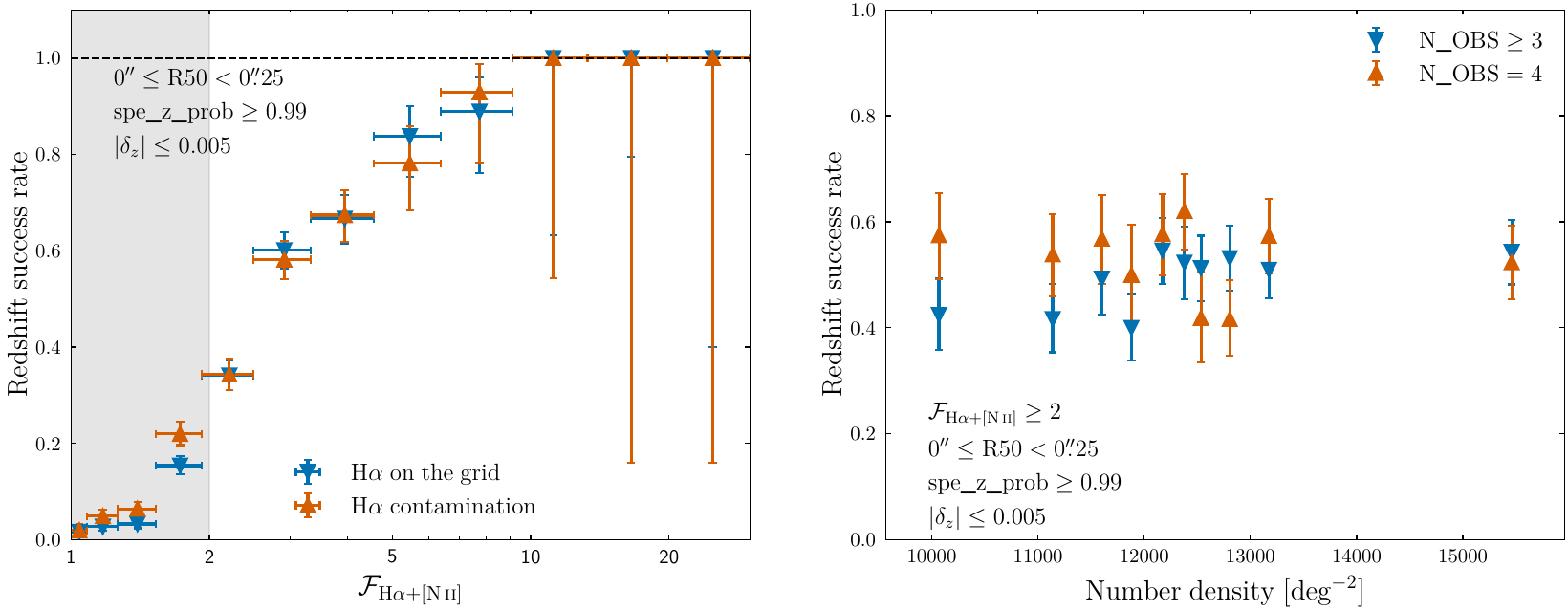}
   \caption{Impact of cross-contamination due to \halpha galaxies only. \emph{Left panel}: success rate as a function of the input \halpha flux. Results for galaxies on the grid are shown in blue, while those with cross-contamination are in orange. \emph{Right panel}: success rate for contamination due to \halpha galaxies as a function of the density in the observed field. The blue markers show spectra observed at least three times across the full wavelength range, while the orange ones correspond to spectra observed four times. Error bars are Clopper--Pearson intervals at 68\% confidence level.}\label{fig:eff_target}
\end{figure*}

Results on the success rate are summarised in Fig. \ref{fig:eff_target}. The left panel shows the success rate as a function of the simulated \halpha + [\ion{N}{ii}] flux for galaxies on the grid and galaxies with contamination from other \halpha galaxies. The sample includes only galaxies with size $\text{R50} < \ang{;;0.25}$. To ensure a fair comparison, the analysis is limited to galaxies observed in the full redshift range four times, as in the case of sources on the grid. Within the error bars, no significant difference is observed between the two cases.

The right panel shows the success rate of the \halpha sample within the 10 observations affected by cross-contamination. Samples include galaxies with flux of the \halpha complex 
$\fhalphacal \geq 2$
. Measurements from spectra observed at least three times and four times in the full wavelength range are reported in blue and orange, respectively. The $x$-axis reports the number density per square degree as in Table \ref{table:obs_coordinates}. We can observe that, given the size of the sample, there is no evidence of a dependence of the redshift success rate on the density of sources.

\subsection{Contamination from all galaxies}
\begin{figure}
    \centering
    \includegraphics[width=\linewidth]{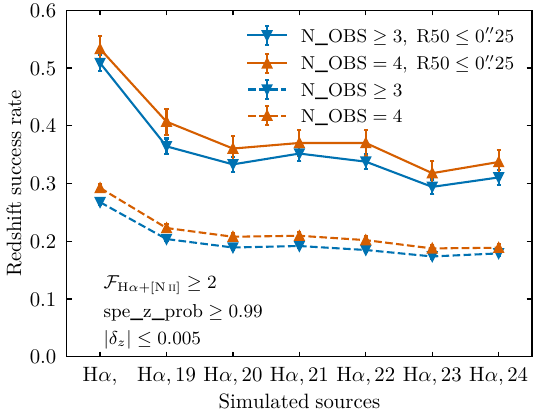}
    \caption{Success rate as a function of the contamination. Orange lines represent spectra observed four times, while blue lines correspond to spectra observed at least three times. Dashed lines show results without any selection on the galaxy size, whereas solid ones are restricted to sources with $\text{R50} < \ang{;;0.25}$. Error bars are Clopper--Pearson intervals at 68\% confidence level.
    }
    \label{fig:sr_contaminants}
\end{figure}

In this section, we evaluate the cross-contamination from galaxies different from \halpha emitters, up to magnitude 24. Figure~\ref{fig:sr_contaminants} shows the success rate of redshift determination as a function of magnitude limit in the simulated objects. The $x$-axis indicates the type of contaminants: `$\mathrm{H}\alpha$' refers to the case discussed in Sect. \ref{sec:5.1}, while `$\mathrm{H}\alpha$, M' indicates contamination from both \halpha galaxies and all galaxies with $\HE\le \rm{M}$. 
The figure displays four different samples. The criteria to determine accurate and reliable measurements, and the minimum flux of the \halpha complex, are the same as in Sect. \ref{sec:5.1}.

As discussed in Sect. \ref{sec:4.2}, the galaxy size has a significant impact on the redshift success rate. Although applying the selection $\text{R50} \leq \ang{;;0.25}$ significantly reduces the sample size, the overall trend of the success rate is similar in all cases. In particular, the success rate drops by approximately 25\% when galaxies brighter than magnitude 19 are added to the \halpha sample, with an additional smaller decrease for contaminants in the [19, 20] magnitude range. Adding sources fainter than 20 mag, the success rate remains constant within the uncertainties. This indicates that, although the number of contaminants increases at fainter magnitudes, their impact on the redshift measurement is limited due to their faint continuum. We can also note that restricting the sample to spectra observed exactly four times, rather than three or more times, results in a gain of 10\% in success rate, at the cost of a reduction in the number of sources of about $\sim 36\%$.

We conclude this section with an estimate of the contamination from galaxies at $0.9 < z <1.8$ that are not included in the \halpha sample, thus violating the assumption that confusion comes from objects that are almost uncorrelated with the \halpha target \Euclid galaxies.
As explained in Sect. \ref{sc:input_catalog}, in our analysis, we remove the clustering information for all the contaminants. Therefore, we will not investigate the contamination effect given by a population of contaminants likely clustered similarly to the \halpha emitters.

\begin{figure}
    \centering
    \includegraphics[width=\linewidth]{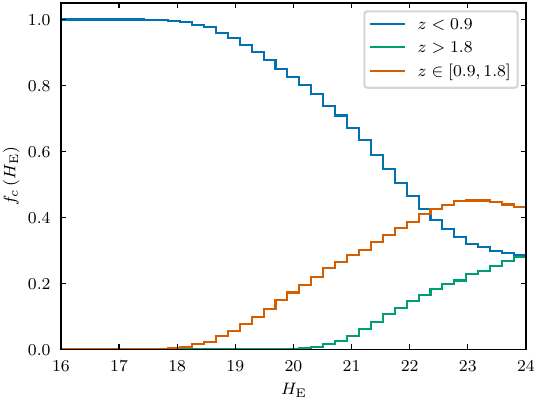}
    \caption{Fraction of contaminant galaxies in a specified redshift range as a function of the magnitude. Contaminants with redshift $z<0.9$ are shown in blue, $z\in [0.9, 1.8]$ in orange, and $z>1.8$ in green. The orange curve is used to estimate the relative loss in success rate in Eq.~\eqref{eq:3} .}
    \label{fig:frac_z}
\end{figure}

We measure the difference in success rate between adjacent magnitude bins as $\Delta \mathcal{SR}_i=\mathcal{SR}(m_i)-\mathcal{SR}(m_{i-1})$, where $m_i$ corresponds to the $i$-th magnitude bin, with $i$ increasing with the magnitude. For an order-of-magnitude estimate, we assume that each contaminant within a given magnitude bin contributes equally to $\Delta \mathcal{SR}_i$. We estimate the relative loss in success rate $L_{\mathcal{SR}, \, z}$ due to contamination of galaxies with redshift $0.9 < z <1.8$, relative to the success rate of contamination within \halpha galaxies only $\mathcal{SR}_{\rm H\alpha}$  as
\begin{equation}
    L_{\mathcal{SR}, \, z} = \sum_i \frac{\Delta \mathcal{SR}_i}{\mathcal{SR}_{\rm H\alpha}}\, f_{c,\,  z\in [0.9, 1.8]}\, (m_i),
    \label{eq:3}
\end{equation}
where $f_{c,\,  z\in [0.9, 1.8]}\, (m_i)$ is the fraction of galaxies in the contaminant sample (as defined in Sect.~\ref{sc:2}) with magnitude in the $i$-th bin and redshift $0.9 < z <1.8$. The fraction $f_{c,\,  z\in [0.9, 1.8]}\, (m_i)$ is estimated from the Flagship catalogue and it is shown in Fig.~\ref{fig:frac_z}. The value of $\Delta \mathcal{SR}_i$ is estimated as the numerical derivative of the curve in Fig.~\ref{fig:sr_contaminants}, which flattens at $\HE \ge 20$ to a very shallow curve with slope $-0.01$ per magnitude. This gives an estimated impact of $L_{\mathcal{SR}, \, z}=$ 4\% due to contaminants in the redshift range $0.9 < z <1.8$. We estimated errors with a Monte Carlo simulation, assuming that the number of successful redshift measurements follows a binomial distribution, and that realizations at different contamination levels are uncorrelated. This method provides an estimated relative error of 50\% for the sample with spectra observed four times, and 34\% for the one observed three times or more (for samples with $\text{R50} \leq \ang{;;0.25}$). However, we note that since our results at different contamination levels are likely correlated (i.e.~we do not have independent samples for each contamination level), this assumption is somehow broken. Having more statistics would allow us to quantify the correlation or, even better, to obtain negligible errors, but at the expense of an excess of computational load. Therefore, the statistical uncertainties we estimate are not accurate, but they likely provide us a useful reference for the order of magnitude of the properly estimated errors.

The limited impact of galaxies within the \halpha redshift range on the redshift success rate is explained by their magnitude distribution. As expected, the observed decrease in success rate is primarily driven by cross-contamination from bright and extended galaxies, which lie at lower redshift compared to the \halpha redshift range. The fraction of galaxies in this range increases with the magnitude and peaks at $\HE\sim 23$ (see Fig.~\ref{fig:frac_z}); at these magnitudes, the continuum is relatively faint, which limits their contribution to the degradation of the redshift success rate.

\section{\label{sc:6}Conclusions}
We presented a new method and analysis based on spectroscopic, pixel-level simulations to assess the performance and potential systematic effects in galaxy redshift measurements within the \gls*{ews}. We extended an existing simulation framework within \Euclid to include the redshift measurement and estimate the impact of cross-contamination induced by galaxies entering in the FOV. 

Simulations used in this study rely on some simplifications, as a complete end-to-end simulation of the photometric and spectroscopic observations demands extensive resources. Given our focus on spectral cross-contamination, we limited our analysis to the spectroscopic channel, using idealised, noiseless photometric cutouts and assuming perfect photometric measurements. Moreover, we only focused on the simulation and measurement of the signal in the first order of the spectra. While this approach may result in an optimistic estimate of the performance, it preserves the validity of our qualitative conclusions regarding cross-contamination.

\bigskip 
We started from a simplified simulation in which \halpha galaxies ($0.9 \le z \le 1.8$) were on purpose simulated on the grid to avoid any cross-contamination. Analysing the distributions of measured redshift, we established a threshold for accurate redshift at $|z_\mathrm{meas}-z_\mathrm{true}| \leq \qty{5e-3}{}$. Under this criterion, 97\% of \halpha galaxies with 
$\fhalphacal \geq 2$
and reliable redshift estimates are accurately measured. We quantified the redshift success rate as a function of both simulated $\mathrm{H\alpha+[\ion{N}{ii}]}$ flux and galaxy size. For compact sources compliant with the \Euclid selection ($\mathrm{R50} < \ang{;;0.25}$)  we found that the success rate goes from 10\% for fluxes 
$\fhalphacal \in [1.4, \, 1.9]$
to 100\% for 
$\fhalphacal \geq 10$
.

\bigskip
We investigated the impact of cross-contamination on the success rate. We started simulating only \halpha galaxies clustered according to the \Euclid Flagship galaxy mock. Results show no significant difference in success rate between clustered \halpha and those on the grid, nor any dependence on the density of the \halpha within the current statistical size of the sample.

Finally, we introduced contaminant galaxies into simulations, increasing their number density by adding progressively fainter sources. We observe a decrease of $\sim 25\%$ in the success rate caused by contaminants of magnitude brighter than 19, with respect to simulations with only \halpha galaxies. The degradation of redshift measurements due to cross-contamination is mainly due to galaxies brighter than magnitude 20; for fainter contaminants, simulations show that the success rate is almost constant within the uncertainties.
We also estimated that the cross-contamination due to galaxies in the \Euclid redshift range $0.9 \leq z \leq 1.8$ reduces the success rate by 4\% with a relative error of order of 50\%.
This is particularly relevant since, at least within the standard cosmological assumptions, the effect of these contaminants is not statistically independent of the cosmological density we are measuring and cannot be mitigated by applying a visibility mask, but requires specific mitigation strategies.
In this respect, we notice that in our analysis, we consider an unclustered population of contaminants, and we therefore assumed clustering will not significantly affect the results. While the impact of this assumption will need to be investigated, for the time being, our results suggest that the small magnitude of this effect is reassuring; determining whether this effect has an impact on the cosmological analyses will require additional investigations, which are beyond the scope of this paper.

\begin{acknowledgements}
\AckEC  
We thank INFN for the support. We acknowledge computational resources by the ReCas HPC-Cluster in Bari and CloudVeneto in Padova.
This work has made use of CosmoHub, developed by PIC (maintained by IFAE and CIEMAT) in collaboration with ICE-CSIC. It received funding from the Spanish government (grant EQC2021-007479-P funded by MCIN/AEI/10.13039/501100011033), the EU NextGeneration/PRTR (PRTR-C17.I1), and the Generalitat de Catalunya. P.M. is supported by the Italian Research Center on High Performance Computing Big Data and Quantum Computing (ICSC) and by the PRIN 2022 PNRR project "Space-based cosmology with \Euclid: the role of High-Performance Computing" (code no. P202259YAF), funded by European Union – Next Generation EU”. F.O., A.R. and C.S. were supported by the MUR PRIN2022 project 20222JBEKN with title "LaScaLa" - funded by the European Union - NextGenerationEU. M.M. acknowledges the financial contribution from the grant PRIN-MUR 2022 2022NY2ZRS 001 ``Optimizing the extraction of cosmological information from Large Scale Structure analysis in view of the next large spectroscopic surveys'' supported by NextGenerationEU, and from the grant ASI n. 2024-10-HH.0 "Attività scientifiche per la missione Euclid – fase E".
We acknowledge the ELSA project.
"ELSA: \Euclid Legacy Science Advanced analysis tools" (Grant Agreement no. 101135203) is
funded by the European Union. Views and opinions expressed are however those of the
author(s) only and do not necessarily reflect those of the European Union or Innovate UK.
Neither the European Union nor the granting authority can be held responsible for them. UK
participation is funded through the UK Horizon guarantee scheme under Innovate UK grant
10093177.

\end{acknowledgements}

\bibliography{my, Euclid}

%

\begin{appendix}
\nolinenumbers

\section{Coordinates of the observations}\label{apx:1}
We selected the sky region to simulate from an area that exhibits a high variance in number density in the Flagship galaxy mock. This region has coordinates $\ang{174}< \mathrm{RA} < \ang{180}$ and $\ang{0} < \mathrm{Dec} < \ang{6}$ and corresponds to an area with high zodiacal background near the ecliptic plane, which will not be observed during the \gls*{ews} \citep{Scaramella-EP1}. Thus, we rotated all the coordinates on the sphere to a region included in the \gls*{ews}. Since the effect of the background light and Milky Way extinction is out of the scope of this paper and can be studied independently with simplified and faster simulations, we chose an optimal region with low background and Milky Way extinction, centred at $\textrm{RA}=\ang{200}$,  $\textrm{Dec}=\ang{60}$.  The selected coordinates of the selected observation are reported in 
Table~\ref{table:obs_coordinates}.

\begin{table}[htbp!]
\smallskip
\caption{Ten simulated observations. For each observation, the table reports the name, the number density of \halpha galaxies computed over NISP footprint, and the coordinates of the centre of the region.}
\label{table:obs_coordinates}
\begin{tabular}{|c|c|c|c|}
\hline
  & & &  \\[-9pt]
  Obs. Id &  Density of \halpha  & RA & Dec\\
  & [$\mathrm{deg}^{-2}$] & [$\mathrm{deg}$] & [$\mathrm{deg}$] \\
  \hline
  & & &  \\[-8pt]
  
F01	& \qty{15459}{}	&200.272928	& 61.183939\\
F02	& \qty{13174}{}	&204.088703	& 61.123210\\
F03	& \qty{12805}{}	&199.715315	& 62.499717\\
F04	& \qty{12534}{}	&203.459208	& 60.350012\\
F05	& \qty{12384}{}	&198.057978	& 61.697008\\
F06	& \qty{12173}{}	&204.818683	& 58.724720\\
F07	& \qty{11880}{}	&199.747877	& 58.552381\\
F08	& \qty{11601}{}	&195.877360	& 61.385861\\
F09	& \qty{11139}{}	&195.144153	& 58.987175\\
F10	& \qty{10068}{}	&195.977509	& 60.597882\\

\hline
\end{tabular}
\end{table}

\section{Bias in the redshift measurement}\label{apx:2}
In this section, we investigate the origin of the redshift estimation bias identified in Fig. \ref{fig:delta_z_grid_vs_z}, which appears independent of the value of the true redshift $z_\text{true}$.  In principle, such bias could arise from two main reasons: a small error in the spectroscopic calibration, or a systematic offset of the centroid of the blended $\mathrm{H\alpha+[\ion{N}{ii}]}$ emission line. In the following, we shall investigate these two possible sources of bias.

\subsection{Spectroscopic calibrations}
The format of the spectroscopic calibrations in the \texttt{SIR} pipeline differs from that used by the \texttt{TIPS} simulator. The \texttt{SIR} pipeline locates the spectra, using a curvature and a dispersion model \citep{Q1-TP006}, both described with Chebyshev polynomials expressed in coordinates of the focal plane and then mapped to pixel coordinates. Instead, \texttt{TIPS} uses \texttt{aXeSIM} \citep{Kummel2009} to simulate spectral traces directly on the focal plane. Since \texttt{aXeSIM} was originally designed for a single detector, its spectroscopic calibrations are expressed using Cartesian polynomials of pixel coordinates. As a result, \texttt{TIPS} adopts a different dispersion law for each detector, whereas \texttt{SIR} uses a global model for the entire focal plane. Therefore, it is not possible to reconstruct the spectra using exactly the same analytical model used in the simulation.

\begin{figure}
    \centering
    \includegraphics[width=\linewidth]{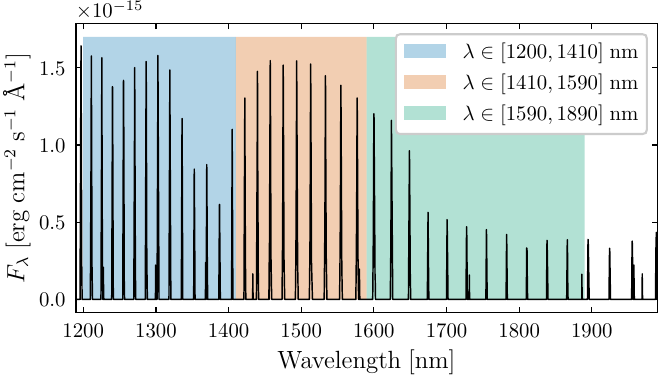}
    \caption{Fabry–P\'erot etalon spectrum from ground-based tests. Colours indicate three different wavelength ranges used to assess the accuracy of the calibrations.}
    \label{fig:spc-etalon}
\end{figure}

To assess the accuracy of the calibrations used in the simulated data processing, we performed a dedicated simulation. We simulated the spectrum of a point-like Fabry–P\'erot etalon used in ground-based tests \citep{EP-Gillard} and parametrised in \cite{gabarra2023}. The spectrum of the etalon contains 33 peaks across the NISP wavelength range, shown in Fig. \ref{fig:spc-etalon}. Sources were simulated on a grid, with $2\times 10$ spectra per detector. 

\begin{figure*}[htbp!]
\centering
    \includegraphics[width=.8\textwidth]{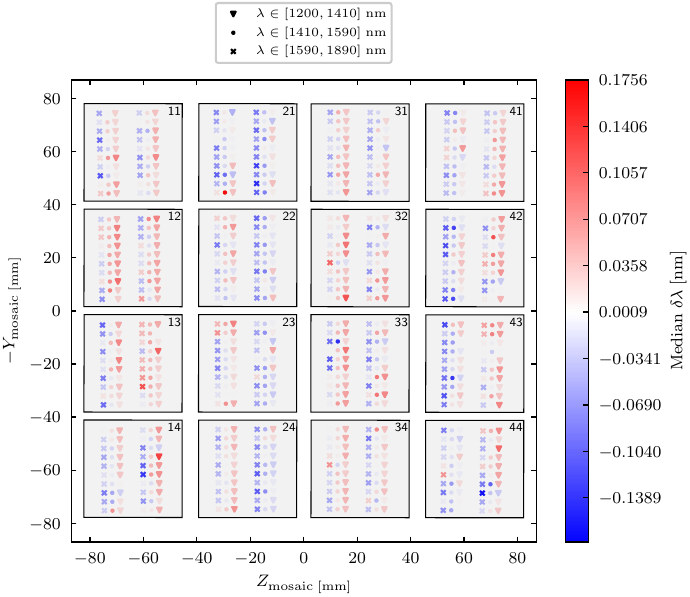}
    \caption{Accuracy of the wavelength calibration in simulated data. Sources are located in $2\times 10$ positions per detector. For each position, markers of different shapes report the median difference between the measured and simulated wavelength in three different wavelength ranges: triangles for  $\lambda \in [1200, 1410]\, \mathrm{nm}$, dots for $\lambda \in [1410, 1590]\, \mathrm{nm}$, and crosses for $\lambda \in [1590, 1890]\, \mathrm{nm}$. The colour bar shows the median difference in each range.}
    \label{fig:spc-calibration}
\end{figure*}

For the analysis, we divided the spectrum into three different regions. For each position, we performed a global fit with 33 Gaussians, excluding the edges of the wavelength range due to their limited sensitivity. Results are summarised in Fig. \ref{fig:spc-calibration}. At each position, the plot shows the median difference between the measured and true centroid of the emission lines for three wavelengths ranges: $\lambda \in [1200, 1410]\,\mathrm{nm}$, $\lambda \in [1410, 1590]\,\mathrm{nm}$, and $\lambda \in [1590, 1890]\,\mathrm{nm}$. These ranges are reported in Fig. \ref{fig:spc-etalon} with different colours.  

The measurements show a good agreement with the simulated wavelengths. On average, smaller wavelengths are slightly underestimated, while the larger ones are overestimated, with a maximum median difference of  $\qty{0.17}{nm}$ from the true value. Thus, the identified bias likely originates from a different effect.

\subsection{Simplified simulations}
We further investigated the origin of the redshift bias with a simplified simulation\footnote{The simulations were performed using the ``ELSA'' tool; a description of the project can be found at \url{https://elsa-euclid.github.io/}.}, in which we did not consider bad pixels, detector inhomogeneities, and cosmic rays. In this simplified framework, the spectra are dispersed using \texttt{SIR} calibration models, ensuring that the same functions are used for both simulation and reconstruction of the spectra. Therefore, this ensures our analysis will not be biased from the different models used for the simulations and spectra extraction.

\begin{figure*}[htbp!]
    \centering
    \includegraphics[width=.6\textwidth]{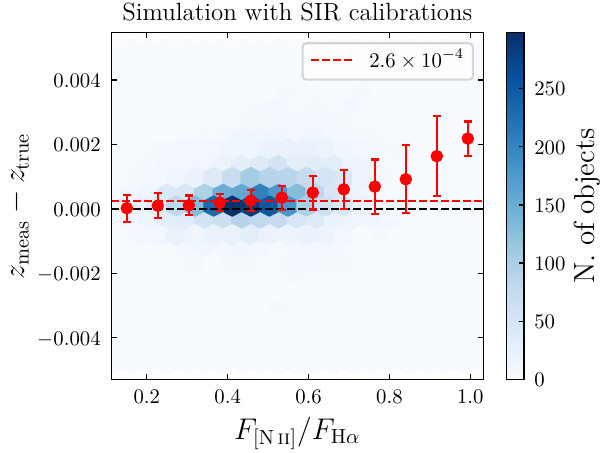}
    \caption{Accuracy of the measured redshift as a function of the [\ion{N}{ii}]/\halpha flux ratio. The red dashed line shows the median difference between the measured and true redshift. Red markers represent the median measured redshift for different [\ion{N}{ii}]/\halpha bins, and error bars show the corresponding MAD.}
    \label{fig:gelsa}
\end{figure*}

Figure \ref{fig:gelsa} shows the difference between the measured and true redshift as a function of the flux ratio between the [\ion{N}{ii}] doublet and the \halpha emission line. The results indicate that the redshift bias increases with the [\ion{N}{ii}]/\halpha flux ratio. This effect arises because, at the NISP spectral resolution, the \halpha line and the [\ion{N}{ii}] doublet are blended. Since the [\ion{N}{ii}] is asymmetric, strong [\ion{N}{ii}] emission leads to offsets in the centroid of the $\mathrm{H\alpha+[\ion{N}{ii}]}$, with a consequence on the measured redshift. Therefore, even if we have not quantified how this bias translates to the bias identified in Fig.~\ref{fig:delta_z_grid_vs_z}, it is likely the main reason of the discrepancy.

\end{appendix}

\end{document}